\documentclass[twocolumn]{aastex701}
\usepackage{amsmath}
\usepackage{xcolor}

\begin{document}

\title{Primitive recovery methods for binary neutron star mergers  with tabulated equations of state in \texttt{SPHINCS\_BSSN}}

\author[orcid=0000-0002-5109-0929]{Swapnil Shankar}
\affiliation{Faculty of Mathematics, Informatics and Natural Sciences, University of Hamburg, Gojenbergsweg 112, 21029 Hamburg, Germany}
\email[show]{swapnil.shankar@uni-hamburg.de}  

\author[orcid=0000-0002-3833-8520]{Stephan Rosswog} 
\affiliation{Faculty of Mathematics, Informatics and Natural Sciences, University of Hamburg, Gojenbergsweg 112, 21029 Hamburg, Germany}
\affiliation{The Oskar Klein Centre, Department of Astronomy, Stockholm University, AlbaNova, SE-10691 Stockholm, Sweden}
\email{stephan.rosswog@uni-hamburg.de}

\author[orcid=0000-0002-3877-0487]{Peter Diener}
\affiliation{Center for Computation \& Technology, Louisiana State University, 70803, Baton Rouge, LA, USA}
\affiliation{Department of Physics \& Astronomy, Louisiana State University, 70803, Baton Rouge, LA, USA}
\email{diener@cct.lsu.edu}

\newcommand{\nifs}{\ensuremath{^{56}\mathrm{Ni}}}
\newcommand{\Nifs}{\ensuremath{^{56}\mathrm{Ni}} $\;$}
\def\paren#1{\left( #1 \right)}
\def\Mesz{M\'esz\'aros~}
\def\Pacz{Paczy\'nski~}
\def\Kluz{Klu\'zniak~}
\def\p{\partial}
\def\msun{M$_{\odot}$}
\def\Msun{M$_{\odot}$ }
\def\be{\begin{equation}}
\def\ee{\end{equation}}
\def\bea{\begin{eqnarray}}
\def\eea{\end{eqnarray}}
\def\bt{\begin{tabbing}}
\def\et{\end{tabbing}}
\def\gcc{g cm$^{-3}$}
\def\Gcc{g cm$^{-3} \;$}
\def\ccm{cm$^3$}
\def\edo{\end{document}}
\def\red{\textcolor{red}}
\def\blue{\textcolor{blue}}
\def\green{\textcolor{green}}
\def\ye{$Y_e$}
\def\Ye{$Y_e$ }
\def\rs{$R_{\rm S}$}
\def\Rs{$R_{\rm S}$ }
\def\noin{\noindent}
\def\QIone{\mathcal{Q}_{\; \; 1}}
\def\QItwo{\mathcal{Q}_{\; \; 2}}
\def\QIthree{\mathcal{Q}_{\; \; 3}}
\def\QIfour{\mathcal{Q}_{\; \; 4}}
\newcommand{\ud}{\mathrm{d}}
\newcommand{\sn}{\ensuremath{\mathrm{sn}}}
\newcommand{\cn}{\ensuremath{\mathrm{cn}}}
\newcommand{\dn}{\ensuremath{\mathrm{dn}}}
\newcommand{\sech}{\ensuremath{\mathrm{sech}}}
\newcommand{\nd}{\noindent}
\newcommand{\enth}{\mathcal{E}}
\newcommand{\carb}{$^{12}$C}
\newcommand{\Carb}{$^{12}$C $\;$}
\newcommand{\ox}{$^{16}$O}
\newcommand{\Ox}{$^{16}$O $\;$}
\newcommand{\tlg}{\tilde{\gamma}}
\newcommand{\tlG}{\tilde{\Gamma}}
\newcommand{\emfp}{e^{-4\phi}}
\newcommand{\tlA}{\tilde{A}}
\newcommand{\tlR}{\tilde{R}}
\newcommand{\xt}{\tilde{x}}
\newcommand{\dt}[1]{\partial_t #1}
\newcommand{\pdu}[3]{\partial_{#3} #1^{#2}}
\newcommand{\pdl}[3]{\partial_{#3} #1_{#2}}
\newcommand{\pdpdu}[4]{\partial_{#3}\partial_{#4} #1^{#2}}
\newcommand{\pdpdl}[4]{\partial_{#3}\partial_{#4} #1_{#2}}
\newcommand{\upwindu}[3]{\beta^{#3}\bar{\partial}_{#3} #1^{#2}}
\newcommand{\upwindl}[3]{\beta^{#3}\bar{\partial}_{#3} #1_{#2}}
\newcommand{\tlGn}{\tilde{\Gamma}_{\mathrm{(n)}}}
\newcommand{\tlGmixed}[2]{\tlG_{#1}^{\;\;\;#2}}

\def\vr{\vec{r}}
\def\ra{\vec{r}_a}
\def\rb{\vec{r}_b}
\def\rab{\vec{r}_{ab}}
\def\rba{(\vec{r}_{ba})}  
\def\rgt{$\vec{r}_{\rm G}$}
\def\rat{$\vec{r}_a$}
\def\rbt{$\vec{r}_b$}
\def\peak{\rm peak}
\def\ji{\varphi}
\def\rin{\mathcal{I}}
\def\x{{\bf x}}
\def\s{{\bf s}}
\def\k{{\bf k}}
\def\n{{\bf n}}
\def\f{{\bf f}}
\def\pd{\partial}
\def\fa{f_{\alpha}(\x,\p,t)}
\def\kabs{\kappa^{\rm abs}_\nu}
\def\ksca{\kappa^{\rm sca}_\nu}
\def\Flux{{\bf F}_\nu}
\def\Frad{{\bf F}_{\rm rad}}
\def\Pressure{{\mathsf P}_\nu}
\def\kB{k_B}
\def\tauph{\tau_{\it ph}}
\def\taumu{\tau_\mu}
\def\MSun{M_{\astrosun}}


\def\M4{M$_4$}
\def\M6{M$_6$}
\def\lcm6{LCM$_6$}
\def\qcm6{QCM$_6$}
\def\whn{W$_{\rm H,n}$}
\def\wh3{W$_{\rm H,3}$}
\def\wh4{W$_{\rm H,4}$}
\def\wh5{W$_{\rm H,5}$}
\def\wh6{W$_{\rm H,6}$}
\def\wh7{W$_{\rm H,7}$}

\def\divv{\nabla \cdot \vec{v}}
\def\vv{\vec{v}}
\def\dx{\Delta x}

\newcommand{\ve}[1]{\boldsymbol{#1}}

\def\ed{\dot{\epsilon}}
\def\vl{v_\lambda}
\def\vort{\vec{\mathcal{W}}}

\newcommand{\rqm}{{%
      \declareslashed{}{\text{-}}{0.04}{0}{I}\slashed{I}}}

\newcommand{\SpB}{\texttt{SPHINCS\_BSSN}\; }
\newcommand{\spB}{\texttt{SPHINCS\_BSSN}}
\newcommand{\spid}{\texttt{SPHINCS\_ID}}
\newcommand{\spiD}{\texttt{SPHINCS\_ID}\; }
\newcommand{\Ma}{\texttt{MAGMA2}\;}
\newcommand{\ma}{\texttt{MAGMA2}}
\newcommand{\Lo}{\texttt{LORENE}\;}
\newcommand{\lo}{\texttt{LORENE}}
\newcommand{\Fu}{\texttt{FUKA}\;}
\newcommand{\fu}{\texttt{FUKA}}
\newcommand{\Sgrid}{\texttt{SGRID \;}}
\newcommand{\sgrid}{\texttt{SGRID}}
\newcommand{\Spells}{\texttt{Spells \;}}
\newcommand{\spells}{\texttt{Spells}}
\newcommand{\Elliptica}{\texttt{Elliptica \;}}
\newcommand{\elliptica}{\texttt{Elliptica}}
\newcommand{\NRPyElliptic}{\texttt{NRPyElliptic \;}}
\newcommand{\nRPyElliptic}{\texttt{NRPyElliptic}}
\newcommand{\comp}{\texttt{CompOSE}\;}
\newcommand{\binns}{\texttt{BinNS}\;}
\newcommand{\McL}{\texttt{McLachlan}\;}
\newcommand{\Mhdu}{\texttt{MHDUET}\;}
\newcommand{\mhdu}{\texttt{MHDUET}}
\newcommand{\Sam}{\texttt{SAMRAI}\;}
\newcommand{\sam}{\texttt{SAMRAI}}
\newcommand{\Simf}{\texttt{Simflowny}\;}
\newcommand{\simf}{\texttt{Simflowny}}

\begin{abstract}
The dynamics and observable signatures of neutron star mergers are 
governed by physics under the most extreme conditions. They are particularly impacted by the high-density equation of state, which for the most sophisticated models is usually available in the form
of tables. Numerical relativity codes usually evolve particularly well-behaved numerical (``conservative") variables, but at the price that the
physically interesting (``primitive") variables need to be found at every computational element and at every integration sub-step by means of expensive (and not always successful) root-finding algorithms. We have recently developed the Lagrangian numerical relativity code \texttt{SPHINCS\_BSSN} which evolves the spacetime on an adaptive mesh with well tested methods, but the fluid is evolved by means of freely moving particles. Since our evolution equations differ from those of conventional numerical relativity, we need to develop new conservative-to-primitive algorithms if we want to use tabulated equations of state. We present here three such algorithms: a 3D and a 2D Newton-Raphson method and a 1D root-finding algorithm based on Ridders' method. We find the 3D method to be very fast and robust with an average failure fraction in a full-blown neutron star merger simulation (with the DD2 equation of state) well below 1\%. While we do not find obvious advantages for the 2D method, the 1D Ridders' method is slow, but essentially fail-safe. Therefore, we choose the 3D Newton-Raphson as default and fall back to the 1D Ridders' method as a safe ``parachute".
\end{abstract}



\section{Introduction}\label{sec:introduction}

The final stages of the inspiral, merger and ringdown of
a neutron star merger are governed by physics at the extremes. These phases
involve strongly curved spacetimes \citep{alcubierre08,baumgarte10,bona09,rezzolla13a,shibata16}, 
potential black hole formation, enormous
magnetic field strengths \citep{price06,anderson08b,rezzolla11,zrake13,kiuchi15,palenzuela22,aguilera22,aguilera24,kiuchi24,aguilera25}, and  densities well beyond nuclear matter density \citep{lattimer12a,lattimer16,baym18,baiotti19}. The post-merger temperatures
 exceed $10^{11}$ K \citep{perego19a}, which results in huge neutrino
luminosities \citep{rosswog03a,dessart09,sekiguchi16a,radice18a,foucart23,schianchi24}. Neutron 
star mergers are further closely related to many grand
challenges of contemporary (astro-)physics including questions related to
dark matter \citep{das21,fortin21,emma22,baryakhtar22}, modifications of general relativity \citep{sakstein17,barack19,jiang21},
cosmic nucleosynthesis \citep{symbalisty82,eichler89,rosswog99,freiburghaus99b,tanvir17,kasen17,arcavi17,rosswog18a,metzger20,cowan21}, cosmic rays \citep{rossoni25,guo25} and gamma-ray bursts \citep{gehrels06,zhang07,troja22,rastinejad22,yang24,levan24}.
Moreover, their ``standard siren" properties \citep{schutz86,holz05,nissanke10} 
make neutron star mergers interesting for an alternative determination of the Hubble parameter.\\
To unlock the full scientific potential of the next generation of ground-based
gravitational wave (GW) detectors such as the Einstein Telescope
or the Cosmic Explorer, theoretical/numerical models need to become more
accurate, both in terms of numerical resolution and in terms of the physical processes that
are modeled.\\
The \SpB \citep{rosswog21a} code is a recent development that combines a general relativistic version of Lagrangian Smoothed Particle Hydrodynamics \citep{monaghan05,rosswog09b,springel10a,price12a,rosswog15c} with spacetime evolution according to the
BSSN equations \citep{alcubierre08,baumgarte10,bona09,rezzolla13a,shibata16} on an adaptive mesh. The recent improvements of the original \SpB methodology
include the use of piecewise polytropic equations of state (EOSs) with an
ideal gas-type thermal part \citep{rosswog22b} and improved mapping of the particle properties
to the adaptive mesh via a local regression estimate (LRE) (e.g. \citet{rosswog2026}). In our most recent
code update \citep{biswas26}, we have, guided by the approach of \cite{etienne24}, also
implemented additional terms in the spacetime evolution to damp constraint violations
and we have implemented (analytical) thermal contributions to internal energy and
pressure that are based on Fermi liquid theory, following the approach of 
\cite{raithel19,raithel21a}.\\
In this paper, our goal is to make tabulated equations of state usable for \spB. This requires the recovery of the physical (``primitive") variables from those variables that are evolved
forward in time, often referred to as ``conservative variables".  This ``conservative-to-primitive"
transformation is a rather non-trivial problem, especially since nuclear matter tables can be
non-smooth, which may require many iterations to convergence, and, in the worst case, one
may even fail to find the correct physical solution. While well-working algorithms 
are available for Eulerian (magneto-)hydrodynamics \citep{cerda_duran08,galeazzi13,Siegel_2018,kastaun21}, these cannot be applied to the
\SpB case, since we have a different set of conservative variables and evolution equations. To make this distinction more concrete: the conservative variables in Eulerian GR(M)HD are the conserved rest mass density $\tilde{D}$, conserved momentum density $\tilde{S}_i$ and conserved energy density $\tilde{\tau}$~\citep{Banyuls_1997, Anton_2006}. In contrast, the conservative variables in \texttt{SPHINCS\_BSSN} are the computing frame density $\rho^*$, specific canonical momentum $S_i$ and specific canonical energy $\hat{e}$. Moreover, while Eulerian GR(M)HD uses Lorentz factor $W$ computed with 3-velocity, \texttt{SPHINCS\_BSSN} uses generalized Lorentz factor $\Theta$ computed with 4-velocity. This means that the algebraic structure of \texttt{con2prim} equations of \texttt{SPHINCS\_BSSN} is very different from Eulerian GR(M)HD, especially for $S_i$ and $\hat{e}$, which means that already existing \texttt{con2prim} methods for Eulerian GR(M)HD cannot be applied without a fundamental rederivation. \\

Here we derive several algorithms to recover the primitive variables for our equation set and we describe and scrutinize their robustness and computational efficiency. The \texttt{con2prim} schemes presented here are a necessary prerequisite for the first \texttt{SPHINCS\_BSSN} binary neutron star merger simulations with a finite-temperature tabulated equation of state, which we demonstrate in section~\ref{sec:BNS_simulation}. The inclusion of essential microphysics, including neutrino physics, for a realistic comparison with gravitational waves and electromagnetic observations, would not be achievable within the \texttt{SPHINCS\_BSSN} framework without the development of these new methods.\\

The outline of the paper is as follows. In section~\ref{sec:3DNR}, \ref{sec:2DNR} and
\ref{sec:1DRidders}, we discuss \texttt{con2prim} schemes that are based on a 3D
Newton-Raphson, a 2D Newton-Raphson and a 1D root finding algorithm according to 
Ridders' method \citep{ridders79,press92}, respectively. We
discuss the results including the convergence properties and costs of all schemes in section~\ref{sec:results}. We then show a proof-of-concept binary neutron star merger
simulation performed with \texttt{SPHINCS\_BSSN} using a tabulated EOS (DD2 EOS) in section~\ref{sec:BNS_simulation} to demonstrate that the newly implemented \texttt{con2prim} schemes are actually very robust in practice. We finally present our conclusions in section~\ref{sec:conclusions}. 

\section{The Lagrangian Numerical Relativity code \spB }
We use geometric units with $G=c=1$ in the equations derived hereafter, and we report the results in the subsequent sections in suitable physical units. We adopt (-,+,+,+) for the 
signature of the metric. We reserve Greek letters for space-time indices from
0...3 with 0 being the temporal component, we use $i$ and $j$ for
spatial components and SPH particles are labeled by $a,b$ and $k$.
The line element and proper time are given by $ds^2= g_{\mu \nu} \; dx^\mu \; 
dx^\nu$ and $d\tau^2= - ds^2$ and the proper time is related to the coordinate
time $t$ by 
\be
\Theta d\tau = dt,
\label{eq:theta_t}
\ee
where we have introduced a generalization of the Lorentz-factor 
\be
\Theta\equiv \frac{1}{\sqrt{-g_{\mu\nu} v^\mu v^\nu}}.
\label{eq:theta}
\ee
The coordinate velocity components are given by
\be
v^\mu= \frac{dx^\mu}{dt}= \frac{dx^\mu}{d\tau} \frac{d\tau}{dt}= \frac{U^\mu}{\Theta}= \frac{U^\mu}{U^0},
\label{eq:v_mu}
\ee
where we have used Eq.~(\ref{eq:theta_t}) and $U^\mu$ is the four-velocity which is normalized to
\be
U^\mu U_\mu= -1. \label{eq:U_norm}
\ee

Thus, the contravariant components of the 3-velocity are related to the 4-velocity as $v^\mu = (1, v^i)$, and the covariant components of the 3-velocity can be calculated using $v_i = g_{i\mu}v^\mu$.
We choose a ``computing frame" (CF) in which the simulation is
performed. In this frame, the mass density, $\rho^\ast$, is calculated 
in a very similar way as the density summation-estimate in Newtonian
SPH\footnote{Note that we no longer use our previous convention of measuring
all energies in units of the baryon rest mass energy $m_0 c^2$. What we used to call CF baryon number density $N$ is now replaced by $\rho^\ast$ and the local rest frame baryon number density $n$ now becomes the density in the local fluid rest frame, $\rho$.}:
\be
\rho_a^\ast= \sum_b m_b W_{ab}(h_a),
\label{eq:rhostar}
\ee
where $m_b$ is the particle mass, $W$ is the SPH smoothing kernel and
$W_{ab}(h_a)= W(|\vec{r}_a - \vec{r}_b|/h_a)$ where $h_a$ is the smoothing length that determines the kernel width. This density is related to the mass density
in the fluid rest frame, $\rho$, via
\be
\rho= \frac{\rho^\ast}{\sqrt{-g} \Theta},
\label{eq:rho_of_rhostar}
\ee
see e.g. \cite{rosswog15c}.
Since our equation set is derived from the Lagrangian of an ideal fluid \citep{monaghan01,rosswog09b,rosswog10a,rosswog15c},  we start from
the canonical momentum $p_i= \frac{\partial L}{\partial v^i}$ 
and the canonical energy $E= \sum_a \frac{\partial L}{\partial v^i_a} v^i_a - L$. We then use the
specific canonical momentum $S_i$ and the specific canonical energy $\hat{e}$ as explained in detail
in Sec.~4.2.3 of \cite{rosswog09b}. This choice
of variables has the advantage that no cumbersome additional time derivatives occur as in older formulations of relativistic SPH \citep{laguna93a,oechslin02} and the 
evolution equations are very similar to the Newtonian case (although for different variables).
Explicitly, the specific canonical momentum reads
\be
S_i= \Theta \mathcal{E} v_i
\label{eq:can_momentum}
\ee
and the specific canonical energy
\be
\hat{e}= S_i v^i + \frac{1 + u}{\Theta} = \Theta \mathcal{E} v_i v^i + \frac{1 + u}{\Theta},
\label{eq:can_energy}
\ee
where $u$ is the specific energy and we have abbreviated the specific enthalpy 
as
\be
\mathcal{E}= 1 + u + \frac{P}{\rho}
\label{eq:enthalpy}
\ee 
with $P$ being the pressure.
The evolution equation for the canonical momentum is then, after some algebra \citep{rosswog09b},
found from the Euler-Lagrange equations as
\be
\frac{d (S_i)_a}{dt} = - \sum_b  m_b 
\left\{
\frac{P_a \; D_i^a}{{\rho_a^{\ast 2}}}   + \frac{P_b \; D_i^b}{\rho_b^{\ast 2}}
\right\}
 + \left(\frac{\sqrt{-g}}{2 \rho^\ast} T^{\mu\nu} 
\frac{\p g_{\mu\nu}}{\p x^i}\right)_a,\label{eq:GR_momentum_evolution}
\ee
where we have used the abbreviations
\be
D^a_i \equiv   \sqrt{-g_a} \;  \frac{\p W_{ab}(h_a)}{\p x_a^i} \quad {\rm and} \quad 
D^b_i \equiv    \sqrt{-g_b} \; \frac{\p W_{ab}(h_b)}{\p x_a^i}
\label{eq:kernel_grad},
\ee
and $T^{\mu \nu}$ is the energy momentum tensor of an ideal fluid

\be
T^{\mu \nu}= \{\rho (1+u) + P\} U^\mu U^\nu + P g^{\mu \nu}.
\ee

The first term in Eq.~(\ref{eq:GR_momentum_evolution}) (involving the summation)
is due to hydrodynamic accelerations and the second term represents the metric
accelerations from the spacetime curvature. In principle,  small corrective terms, so-called ``grad-h" terms, also appear in the general relativistic version of 
SPH~\citep{rosswog10a}. Their main purpose is to further improve the (already very good) overall conservation. But their effect has been found to be very minor in a direct comparison \citep{rosswog07c} and 
their calculation requires an additional iteration between densities and smoothing lengths. We therefore ignore these terms here.
The evolution equation of canonical energy 
follows from straight forwardly taking the Lagrangian time derivative of Eq.~(\ref{eq:can_energy}), applying the first law of thermodynamics and after some algebra
\citep{rosswog10a} one obtains

\be
\frac{d\hat{e}_a}{dt}= - \sum_b m_b \left\{
\frac{P_a v^i_b}{\rho_a^{\ast 2}} \; D_i^a  + 
\frac{P_b v^i_a}{\rho_b^{\ast 2}} \; D_i^b   
\right\}
 - \left( \frac{\sqrt{-g}}{2 \rho^\ast} T^{\mu\nu} \p_t g_{\mu\nu}\right)_a.\label{eq:GR_energy_evolution}
\ee
To robustly capture shocks, the equations are augmented by
modern shock dissipation terms. Since their description is rather technical and lengthy, we refer the interested reader
to Sec.~2.1.1 of \cite{rosswog22b}, where these terms are explained in detail.\\

The evolution equations (\ref{eq:GR_momentum_evolution}) and (\ref{eq:GR_energy_evolution}) depend 
on the ``primitive variables" mass density ($\rho$), the specific energy density ($u$) and the 3-velocity ($v^i$) or quantities that are derived from them
(e.g. the pressure $P$). For the variables that
we really evolve forward in time, namely $\hat{e}$ and $S_i$, and $\rho^\ast$ (which
is calculated each time step from the current positions and smoothing lengths), we 
use the term ``conservative variables" to stay consistent with the language used 
in a Eulerian context. The primitive variables need to be recovered from the conservative ones at every Runge-Kutta sub-step. It is straight forward
to calculate the conservative variables from
the primitive ones, hereafter called \texttt{prim2con},
see Eqs.~(\ref{eq:rhostar}), (\ref{eq:can_momentum}) and (\ref{eq:can_energy}), but recovering the primitive
variables from the conservative ones (hereafter \texttt{con2prim}) is a 
rather intricate process which requires nested loops of numerical 
root-finding.

\begin{figure*}
\centering
\includegraphics[width=1.0\textwidth]{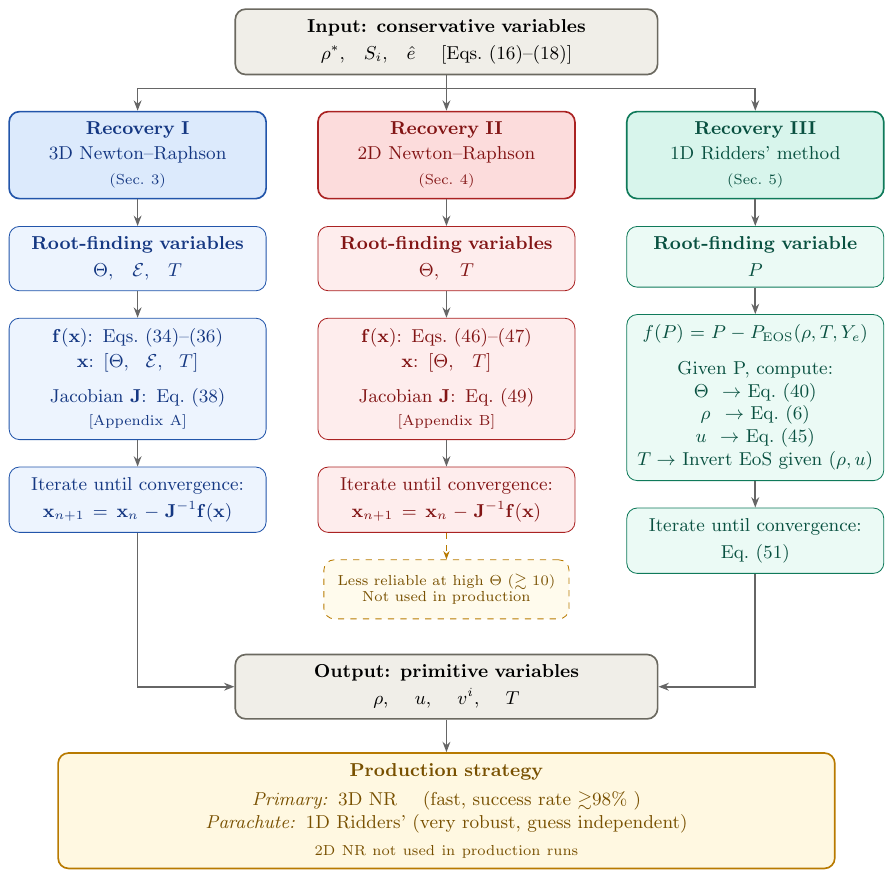}
\caption{Summary of the recovery methods}
\label{fig:flowchart_recovery}
\end{figure*}

Our method to perform \texttt{con2prim} for simple polytropes is explained in \citep{rosswog21a} and the extension to piecewise polytropic EOSs with an ideal thermal contribution is shown in detail in Appendix A of \citep{rosswog22b}. For the first algorithm we had used a straight forward Newton-Raphson scheme, for the second one
we used Ridders' method \citep{ridders79,press92} which has turned out to be very fast and robust.
In this work, we focus on \texttt{con2prim} schemes in the context of equations of state that come in a tabulated form (and hence are also called \textit{tabulated} EOSs). For a tabulated EOS, thermodynamic variables such as pressure, specific energy density, entropy etc. are provided in a tabulated form as a discrete function of density $\rho$, temperature $T$ and electron fraction $Y_e$. One therefore needs to perform interpolations to calculate these thermodynamic variables at any given ($\rho$, $T$, $Y_e$).\\
We suggest in the following sections three different recovery approaches:
\begin{itemize}
\item 3D Newton-Raphson: this approach is rather straight forward, but works fast and reliably in nearly 100\% of the cases.
\item 2D Newton-Raphson: this lower-dimensional root-finding scheme works well, apart from extreme cases, but it turns out, see Sec.~\ref{sec:results}, that it has no obvious advantage in comparison with the other two methods.
\item 1D root finding with Ridders' method: while this algorithm takes by far the most EOS calls, it turns out to be essentially fail-safe and it is therefore an ideal ``parachute" method for the rare cases where the faster methods fail.
\end{itemize}
We summarize these three recovery methods in the flowcharts shown in  Fig.~\ref{fig:flowchart_recovery}.

\section{Recovery I: 3D Newton-Raphson}\label{sec:3DNR}
In this section, we describe how the Newton-Raphson method can be used to iteratively find the primitives ($\rho$, $u$, $v^x$, $v^y$, $v^z$) given a set of conservatives ($\rho^*$, $S_i$, $\hat{e}$). This method is designed to be used with \textit{tabulated EOS}. 

\subsection{Introduction to Newton-Raphson}
Given a real-valued function $f(x)$ and an initial guess $x_0$, one can approximate a root of $f(x)$ by iteratively improving a guess $x_n$ via
\begin{equation}
    x_{n+1} = x_n - \frac{f(x)}{f'(x)},
\end{equation}
where  $f'(x)$ is the derivative of $f(x)$ with respect to $x$. This formula can be generalized to a system of $M$ equations in $N$ variables as follows

\begin{equation}\label{eq:NR-iterative-step}
    \mathbf{x}_{n+1} = \mathbf{x}_{n} - \mathbf{J}^{-1} \mathbf{f}(\mathbf{x}) = \mathbf{x}_{n} + \mathbf{\delta x}_{n}, 
\end{equation}

where $\mathbf{x} = [x^1, x^2, ..., x^N]^T$ is the vector of variables to be solved for, $\mathbf{f}(\mathbf{x}) = [f^1(\mathbf{x}), f^2(\mathbf{x}), ...., f^M(\mathbf{x})]^T$ is the vector of real-valued functions whose roots we need to find and $\mathbf{J}$ is the Jacobian matrix defined as $J_{ij} = \partial{f^i}/\partial{x^j}$.

\subsection{Reduced system of equations for \texttt{con2prim}}
Our equation system
\begin{eqnarray} 
\rho^* &=& \sqrt{-g}\,\Theta\,\rho \label{eq:p2c1} \\ 
S_i &=& \Theta \, \mathcal{E} \, v_i \label{eq:p2c2} \\ 
\hat{e} &=& S_i v^i + \frac{1+u}{\Theta}\label{eq:p2c3} 
\end{eqnarray}
forms a system of 5 algebraic equations in 5 unknowns ($\rho$, $u$, $v^x$, $v^y$, $v^z$). To this end, we first reduce the number of equations and unknowns to 3 by making use of certain scalars as shown in the following paragraphs. We then cast the equations in a form such that $v^i$ does not appear explicitly in any equation. We aim to use the generalized Lorentz factor $\Theta$, specific enthalpy $\mathcal{E}$ and temperature $T$ as the root-finding variables for our Newton-Raphson method. 

We first derive some useful relations for later use.
Multiplying  $S_i$ with $v^i$ yields
\begin{equation} \label{eq:Sivi}
    S_i v^i = \Theta \mathcal{E} v_i v^i 
\end{equation}
and we find
\begin{eqnarray}
S^2 = S_i S^i &=& g^{ij} S_i S_j
= \Theta^2 \mathcal{E}^2 v_i (g^{ij} v_j).   
\end{eqnarray}
With $v^i = g^{i\mu} v_\mu = g^{i0} v_0 + g^{ij} v_j$ $\Rightarrow g^{ij} v_j = v^i - g^{i0} v_0$,
therefore,
\begin{eqnarray}
S^2 &=& \Theta^2 \mathcal{E}^2 v_i (v^i - g^{i0} v_0) \nonumber \\
&=& \Theta^2 \mathcal{E}^2 v_i v^i - \Theta^2 \mathcal{E}^2 v_0 g^{i0} v_i \nonumber \\
&=& \Theta^2 \mathcal{E}^2 v_i v^i - \Theta \mathcal{E} v_0 g^{i0} S_i \nonumber   
\end{eqnarray}
or,
\begin{equation} \label{eq:SiSi}
    S^2 + \Theta \mathcal{E}  v_0 g^{0j} S_j = \Theta^2 \mathcal{E}^2 v_i v^i.
\end{equation}

\subsubsection{First reduced equation}
Combining Eqs.~(\ref{eq:Sivi}) and (\ref{eq:SiSi}), 
and solving for $S_i v^i$ we find
\begin{equation} \label{eq:Sivi_final}
 S_i v^i = \frac{S^2}{\Theta \mathcal{E}} +  v_0 g^{0j} S_j .
\end{equation}

Substituting $S_iv^i$ from Eq.~(\ref{eq:Sivi_final}) in Eq.~(\ref{eq:p2c3}), we get
\begin{eqnarray}
    \hat{e} 
    &=& \frac{S^2}{\Theta\mathcal{E}} + v_0 g^{0j} S_j + \frac{1+u}{\Theta} \\
    &=& \frac{S^2}{\Theta\mathcal{E}} + v_0 g^{0j} S_j +  \frac{1}{\Theta}\left( \mathcal{E} - \frac{P\Theta \sqrt{-g}}{\rho^*}\right), 
\end{eqnarray}
where we have used Eqs.~(\ref{eq:rho_of_rhostar}) and (\ref{eq:enthalpy}).
The covariant time component of the coordinate velocity 
can be written as
\be
v^0 = 1 = g^{0\mu} v_\mu = g^{00} v_0 + g^{0i} v_i \\
    \Rightarrow v_0 = \frac{1-g^{0i} v_i}{g^{00}}.
    \label{eq:v_0v2}
\ee
Substituting $v_0$, using Eq.~(\ref{eq:p2c2}) and rearranging the terms, we finally get the first reduced equation 
\begin{multline}
\Theta \mathcal{E} \rho^* \hat{e}  - \rho^* S^2 - \mathcal{E}^2 \rho^* + \Theta \mathcal{E} \sqrt{-g}\, P_\text{EOS} \\
- \frac{g^{0j}S_j}{g^{00}} \left[ \Theta \mathcal{E} \rho^* - \rho^* g^{0j} S_j \right] = 0.
\end{multline}
Note that 
\begin{equation}
P_{\text{EOS}} = P_{\text{EOS}}(\rho,T,Y_e) \label{eq:Peos}
\end{equation}
is calculated from the tabulated EOS, where $\rho$, see Eq.~(\ref{eq:rho_of_rhostar}), is only a function of the 
root-finding and the conservative variables.

\subsubsection{Second reduced equation}
To find the next reduced equation, we use the definition of generalized Lorentz factor as follows
\begin{equation} \label{eq:theta_expansion}
    -\Theta^{-2} = g_{\mu\nu} v^\mu v^\nu = v^\mu v_\mu = v^0 v_0 + v^i v_i = v_0 + v^i v_i.  
\end{equation}
Substituting $v_i$ from Eq.~(\ref{eq:p2c2}) in Eq.~(\ref{eq:v_0v2}) and using it
 in Eq.~(\ref{eq:theta_expansion}), we find
\begin{equation} \label{eq:vivi}
    v^iv_i = -v_0 - \frac{1}{\Theta^2} = \frac{1}{g^{00}} \left( \frac{g^{0j}S_j}{\Theta\mathcal{E}} - 1 \right) - \frac{1}{\Theta^2}.
\end{equation}
Finally, substituting $v^iv_i$ from Eq.~(\ref{eq:vivi}) in Eq.~(\ref{eq:p2c3}), we get
\begin{eqnarray} 
\hat{e} &=& \Theta \mathcal{E} v^i v_i + \frac{1+u}{\Theta} \nonumber \\
&=& \Theta \mathcal{E} \left\{ \frac{1}{g^{00}} \left( \frac{g^{0j}S_j}{\Theta\mathcal{E}} - 1 \right) - \frac{1}{\Theta^2} \right\} + \frac{1+u}{\Theta} \nonumber \\ 
&=& \frac{g^{0j}S_j}{g^{00}} - \frac{\Theta\mathcal{E}}{g^{00}} - \frac{\mathcal{E}}{\Theta}  + \frac{1+u}{\Theta} \nonumber \\ 
&=& \frac{g^{0j}S_j}{g^{00}} - \frac{\Theta\mathcal{E}}{g^{00}} - \frac{P}{\Theta \rho}  \nonumber \\ 
&=& \frac{g^{0j}S_j}{g^{00}} - \frac{\Theta\mathcal{E}}{g^{00}} - \frac{P \sqrt{-g}}{\rho^*}.  
\end{eqnarray}
Rearranging the terms, we obtain the  second reduced equation as
\begin{equation}
  \rho^* \Theta \mathcal{E} - \rho^* g^{0j} S_j + g^{00}P_{\text{EOS}}\sqrt{-g} + \rho^* g^{00} \hat{e} = 0.    
\end{equation}
As for the first reduced equation, 
$P_{\text{EOS}} = P_{\text{EOS}}(\rho,T,Y_e)$ is calculated from the tabulated EOS with $\rho=\frac{\rho^*}{\Theta\sqrt{-g}}$, so the equation 
only depends on the root-finding variables and conservatives.

\subsubsection{Third reduced equation}
We use the equation of specific energy $u$ directly as the third reduced equation in order to use the temperature $T$ as one of the root-finding variables. The equation is as follows
\begin{equation}
    u - u_{\text{EOS}} = 0,
\end{equation}
where $u_{\text{EOS}} = u_{\text{EOS}}(\rho,T,Y_e)$ is calculated from the tabulated EOS as before and $u$ is determined using
\begin{equation}
    u = \mathcal{E} - 1 - \frac{P}{\rho} = \mathcal{E} -1 - \frac{P_{\text{EOS}}\Theta\sqrt{-g}}{\rho^*},
\end{equation}
which is again a function of only the root-finding variables ($\Theta$, $\mathcal{E}$, $T$) and conservatives.

\subsection{Final equations}\label{sec:3DNR-final-equations}
We thus have the following  three equations
\begin{equation}\label{eq:3DNR_final1}
\begin{aligned}
&\Theta \mathcal{E} \rho^* \hat{e}  - \rho^* S^2 - \mathcal{E}^2 \rho^* + \Theta \mathcal{E} \sqrt{-g}\, P_\text{EOS} \\
&\qquad - \frac{g^{0j}S_j}{g^{00}} \left[ \Theta \mathcal{E} \rho^* - \rho^* g^{0j} S_j \right] = 0
\end{aligned}
\end{equation}
\begin{equation}\label{eq:3DNR_final2}
\begin{aligned}
&\rho^* \Theta \mathcal{E} - \rho^* g^{0j} S_j + g^{00}P_{\text{EOS}}\sqrt{-g} \\
&\qquad + \rho^* g^{00} \hat{e} = 0
\end{aligned}
\end{equation}
\begin{equation}\label{eq:3DNR_final3}
u - u_{\text{EOS}} = 0
\end{equation}
that we solve by means of a 3D Newton-Raphson method with variables $\Theta$, $\mathcal{E}$ and $T$, where
the vector of unknowns $\mathbf{x} = [\Theta, \mathcal{E}, T]^T$ and $\delta \mathbf{x} = [\delta \Theta, \delta \mathcal{E}, \delta T]^T$.  The function $\mathbf{f}(\mathbf{x})$ is given by
\begin{eqnarray}
    \mathbf{f}(\mathbf{x}) &=& 
    \left[\begin{array}{c}
          f^1(\Theta, \mathcal{E}, T) \\
          f^2(\Theta, \mathcal{E}, T) \\
          f^3(\Theta, \mathcal{E}, T) 
    \end{array}\right] =  \left[\begin{array}{c}
          \text{LHS of Eq.~(\ref{eq:3DNR_final1})} \\
          \text{LHS of Eq.~(\ref{eq:3DNR_final2})} \\
          \text{LHS of Eq.~(\ref{eq:3DNR_final3})} 
    \end{array}\right] \label{eq:3DNR-fx} 
\end{eqnarray}
and the Jacobian matrix is given by
\begin{eqnarray}
\mathbf{J} = 
\left[\begin{array}{ccc}
\frac{\partial f^1}{\partial\Theta} & \frac{\partial f^1}{\partial\mathcal{E}} & \frac{\partial f^1}{\partial T} \\
\frac{\partial f^2}{\partial\Theta} & \frac{\partial f^2}{\partial\mathcal{E}} & \frac{\partial f^2}{\partial T} \\
\frac{\partial f^3}{\partial\Theta} & \frac{\partial f^3}{\partial\mathcal{E}} & \frac{\partial f^3}{\partial T} 
\end{array}\right].
\label{eq:Jacobian_3DNR}
\end{eqnarray}
We explicitly show how to calculate these derivatives in Appendix~\ref{appendix:jacobian_3DNR}.\\
After each iterative update of $\mathbf{x}$ according to Eq.~(\ref{eq:NR-iterative-step}), we bound the temperature $\mathbf{x}[3]$ between the minimum and maximum of the EOS table. We calculate the recovery error $\varepsilon$, defined as
\begin{equation}\label{eq:error-3DNR}
    \varepsilon = \mathrm{max}\left[\left|\frac{\Theta_{n+1}-\Theta_n}{\Theta_n} \right|, \left|\frac{\mathcal{E}_{n+1}-\mathcal{E}_{n}}{\mathcal{E}_{n}}\right|,\left|\frac{T_{n+1}-T_n}{T_n}\right|\right],
\end{equation}
at each iteration.  Convergence is reached when the recovery error falls below our chosen tolerance of $10^{-10}$. If the recovery error does not fall below the tolerance after $128$ 
iterations, we determine the Newton-Raphson method has failed to converge and thus \texttt{con2prim} has failed. If the iterations converge, we calculate $\rho$ by substituting the new $\Theta$ in Eq.~(\ref{eq:p2c1}), and then use the new $\rho$ and $T$ to determine $P$ and $u$ from the EOS table. We determine $v_0$ using Eq.~(\ref{eq:vivi}) and $v_i$ using Eq.~(\ref{eq:p2c2}), and then calculate the new velocity $v^i$ using $v^i = g^{i\nu}v_\nu$.

\section{Recovery II: 2D Newton-Raphson}\label{sec:2DNR}

\subsection{Reduced system of equations for \texttt{con2prim}}
In this case, we reduce the number of equations and unknowns to two. We aim to use the generalized Lorentz factor $\Theta$ and the temperature $T$ as the root-finding variables for our Newton-Raphson method.

\subsubsection{First reduced equation}
The generalized Lorentz factor $\Theta$ can be expressed in terms of pressure, see Eq.~(51) in \cite{rosswog21a}, as follows:
\begin{equation} \label{eq:theta_given_P}
    \Theta = \sqrt{ \frac{-g^{00}}{1 + \frac{A}{B^2}} },
\end{equation}
where
\begin{equation}
    A = g^{00}g^{jk}S_jS_k - (g^{0j}S_j)^2
\end{equation}
and
\begin{equation}
    B = B(P) = g^{0j}S_j - g^{00}\left( \frac{\sqrt{-g}}{\rho^*} P + \hat{e} \right).
\end{equation}
Simplifying, we find the first reduced equation:
\begin{equation}
    \Theta^2 ( B^2 + A) + g^{00} B^2 = 0,
\end{equation}
where it should be noted that $B$ is a function of pressure $P_{\text{EOS}}$, see Eq.~(\ref{eq:Peos}).

\subsubsection{Second reduced equation}
We again use the equation of the specific energy $u$ directly, as in the third reduced equation of 3D Newton-Raphson, in order to use the temperature $T$ as one of the root-finding variables. The equation is as follows
\begin{equation}
    u - u_{\text{EOS}} = 0
\end{equation}
The specific energy $u$ can be expressed in terms of $\Theta$ and $P$, see Eq.~(49) in \cite{rosswog21a}, as follows:
\begin{equation} \label{eq:u_given_theta_P}
    u = \frac{1}{\Theta} ( g^{0j}S_j - g^{00}\hat{e}) - \frac{\sqrt{-g}P_{\text{EOS}}}{\Theta \rho^*}(g^{00} + \Theta^2) - 1,
\end{equation}
which is a function of root-finding variables ($\Theta$, $T$) and conservatives. 

\subsubsection{Final equations}\label{sec:2DNR-final-equations}
Our final equation set then reads 
\begin{eqnarray}
\Theta^2 ( B^2 + A) + g^{00} B^2 &=& 0 \\
u - u_{\text{EOS}} &=& 0.
\end{eqnarray}
We solve this system of equations with a 2D Newton-Raphson method with variables $\Theta$ and $T$. Thus, the vector of unknowns $\mathbf{x} = [\Theta, T]^T$ and $\delta \mathbf{x} = [\delta \Theta, \delta T]^T$.  The function $\mathbf{f}(\mathbf{x})$ is given by
\begin{eqnarray}
    \mathbf{f}(\mathbf{x}) &=& 
    \left[\begin{array}{c}
          f^1(\Theta, T) \\
          f^2(\Theta, T)  
    \end{array}\right] = 
    \left[\begin{array}{c} 
          \Theta^2 ( B^2 + A) + g^{00} B^2  \\ 
          u - u_{\text{EOS}} \\
    \end{array} \right]   \label{eq:2DNR-fx} 
\end{eqnarray}
and the Jacobian matrix  by
\begin{eqnarray}
\mathbf{J} = 
\left[\begin{array}{ccc}
\frac{\partial f^1}{\partial\Theta}  & \frac{\partial f^1}{\partial T} \\
\frac{\partial f^2}{\partial\Theta}  & \frac{\partial f^2}{\partial T} 
\end{array}\right]
\label{eq:Jacobian_2DNR}
\end{eqnarray}
We explicitly show how to calculate these derivatives in Appendix~\ref{appendix:jacobian_2DNR}.\\
We use the same iterative approach as Sec~\ref{sec:3DNR-final-equations} with a modified definition of recovery error $\varepsilon$ given by
\begin{equation}\label{eq:error-2DNR}
    \varepsilon = \mathrm{max}\left[\left|\frac{\Theta_{n+1}-\Theta_n}{\Theta_n}\right|, \left|\frac{T_{n+1}-T_{n}}{T_{n}}\right|\right],
\end{equation}

Similar to the 3D Newton-Raphson method, we again use the recovery error tolerance of $10^{-10}$ to determine convergence and use a maximum of $128$ iterations. If the iterations converge, we determine $\rho$ by substituting the new $\Theta$ in Eq.~(\ref{eq:p2c1}). We then use the new $\rho$ and $T$ to calculate $P$ and $u$ from the EOS table. We then substitute the new $P$, $u$ and $\rho$ in Eq.~(\ref{eq:enthalpy}) to calculate $\mathcal{E}$. Since the new $\Theta$ and $\mathcal{E}$ are now known, we can calculate $v^i$ using the same procedure as described in Sec~\ref{sec:3DNR-final-equations}.

\section{Recovery III: robust 1D root finding with Ridders' method}\label{sec:1DRidders}
\subsection{Introduction to Ridders' method for root finding}
A powerful root finding method is due to \cite{ridders79}.
Given a real-valued non-linear function $f(x)$ such that a root of the function is known to lie within the interval $[a,b]$, i.e. $f(a) f(b) < 0$, one can find a new point $c$, which is closer to the root than $a$ or $b$ as follows
\begin{equation}
    c = m + (m-a)\,\mathrm{sign}\!\big[f(a)-f(b)\big]\,\frac{f(m)}{\sqrt{ f(m)^2 - f(a) f(b) }},
\end{equation}
where $m = \frac{1}{2}(a+b)$ is the midpoint of the current bracket. If $f(c) > 0$, then the new bracket becomes $[a,c]$, else the new bracket becomes $[c,b]$. This method combines the robustness of bracketing methods such as bisection, with quadratic convergence per iteration, similar to Newton's method. However, this method requires the function evaluation twice at every iteration (at $m$ and $c$), and thus the effective order of convergence with respect to function evaluations is $\sqrt{2}$ i.e. superlinear convergence. Unlike Newton-Raphson, this method does not require any derivatives, which sometimes are 1) not available, 2) costly to calculate or, 3) if obtained by finite differencing, may not be very accurate and therefore can lead to slower convergence or failure. 

\subsection{Methodology}
We use the pressure $P$ as the root-finding variable with $f(P)$ given by
\begin{equation}
    f(P) = P - P_\text{EOS}(\rho, T, Y_e),
\end{equation}
where we need to determine $\rho$ and $T$ using $P$ at a given iteration step in Ridders' method. In order to determine $\rho$ and $T$, and subsequently $f(P)$, from the pressure $P$ at the current iteration, we perform the following steps:
\begin{enumerate}
    \item We first determine the generalized Lorentz factor $\Theta$ using $P$ and the conservatives with Eq.~(\ref{eq:theta_given_P}).
    \item Once $\Theta$ is known, we can determine $\rho$ using Eq.~(\ref{eq:rho_of_rhostar}).
    \item We then determine the specific energy $u$ using $P$ and $\Theta$ with Eq.~(\ref{eq:u_given_theta_P}). After the determination of $u$, we also find the minimum and maximum value of $u$ permitted within the EOS table, $u_{\mathrm{min}}(\rho, Y_e)$ and $u_{\mathrm{max}}(\rho, Y_e)$, for the given $\rho$ and  $Y_e$. We then limit $u$ between $u_{\mathrm{min}}$ and $u_{\mathrm{max}}$. This step is important in order to ensure that a temperature inversion of $u$ is always possible, which we describe in the next step.
    \item We now know $\rho$ and $u$ at a given iteration step. The equation of state table provides $u(\rho,T,Y_e)$, which means that we need to invert the table in order to find the temperature $T$ for a given $\rho$, $Y_e$ and $u$. This requires another root-finding iterative method, for which we again use the Ridders' method. This additional root-finding for $T$ at every iteration makes this method substantially more expensive compared to 2D or 3D methods, where $T$ is also a root-finding variable which makes this temperature inversion unnecessary. 
\end{enumerate}
Once we know $\rho$ and $T$ for a given value of $P$, we can determine the Ridders' function $f(P)$. We calculate the recovery error $\varepsilon$ at each iteration using
\begin{equation}\label{eq:error-1DRidders}
    \varepsilon = \left|\frac{P_{n+1}-P_n}{P_n}\right| 
\end{equation}

 We determine that the method has converged when the recovery error falls below our chosen tolerance of $10^{-10}$ and the condition $|f(P_n)| \leq 10^{-10} |P_n|$ is satisfied. We perform a maximum of $100$ iterations after which we determine that \texttt{con2prim} has failed. Once the iterations have converged, we use the same steps as described above to calculate the new $\Theta$, $\rho$, $u$ and $T$ using the new converged $P$. We then use the new $\rho$, $u$ and $P$ to calculate $\mathcal{E}$, and then determine the new velocity $v^i$ using the new $\Theta$ and $\mathcal{E}$ using the approach described in Sec.~\ref{sec:2DNR-final-equations}.

In practice, we start with the pressure bracket $[P_{\text{min}}, P_{\text{max}}]$, where $P_{\text{min}}$ and $P_{\text{max}}$ are the minimum and maximum pressure in EOS table. This choice of bracket allows us to guarantee that a root exists a priori, with a couple of caveats. The pressure $P_\text{EOS}$ is obtained from the table by trilinear interpolation, and therefore cannot lie outside the range of the table i.e. $P_{\text{min}} \leq P_{\text{EOS}} \leq P_{\text{max}}$ for any input $(\rho, T, Y_e)$ within the table bounds. Consequently, $f(P_{\text{min}}) = P_{\text{min}} - P_{\text{EOS}} \le 0$ and $f(P_{\text{max}}) = P_{\text{max}} - P_{\text{EOS}} \ge 0$, which means $f$ changes sign across our chosen initial  bracket. Since Ridders' method is a bracketing scheme, the root remains enclosed at every iteration and convergence is guaranteed. However, this argument requires $f$ to be well-defined throughout the iterative steps which can fail if $\rho$ goes out of the table bounds during any iterative step as in step 2 above, or the iterations fail to converge within the desired tolerance due to limiting $u$ between table bounds repeatedly as in step 3 above. Both these failure modes can be triggered at table edges, which is the reason we observe some failures in the edge cases as seen in Fig.~\ref{fig:convergence_all}. Our choice of initial pressure bracket from the EOS table bounds means that Ridders' method for \texttt{con2prim} is {\em independent of the initial guess}, unlike the Newton-Raphson method which depends on the initial guess for primitives. This further improves the robustness of Ridders' method which we will demonstrate in Sec.~\ref{sec:results}.

\section{Results}\label{sec:results}

\begin{figure*}
\centering
\includegraphics[width=1.0\textwidth]{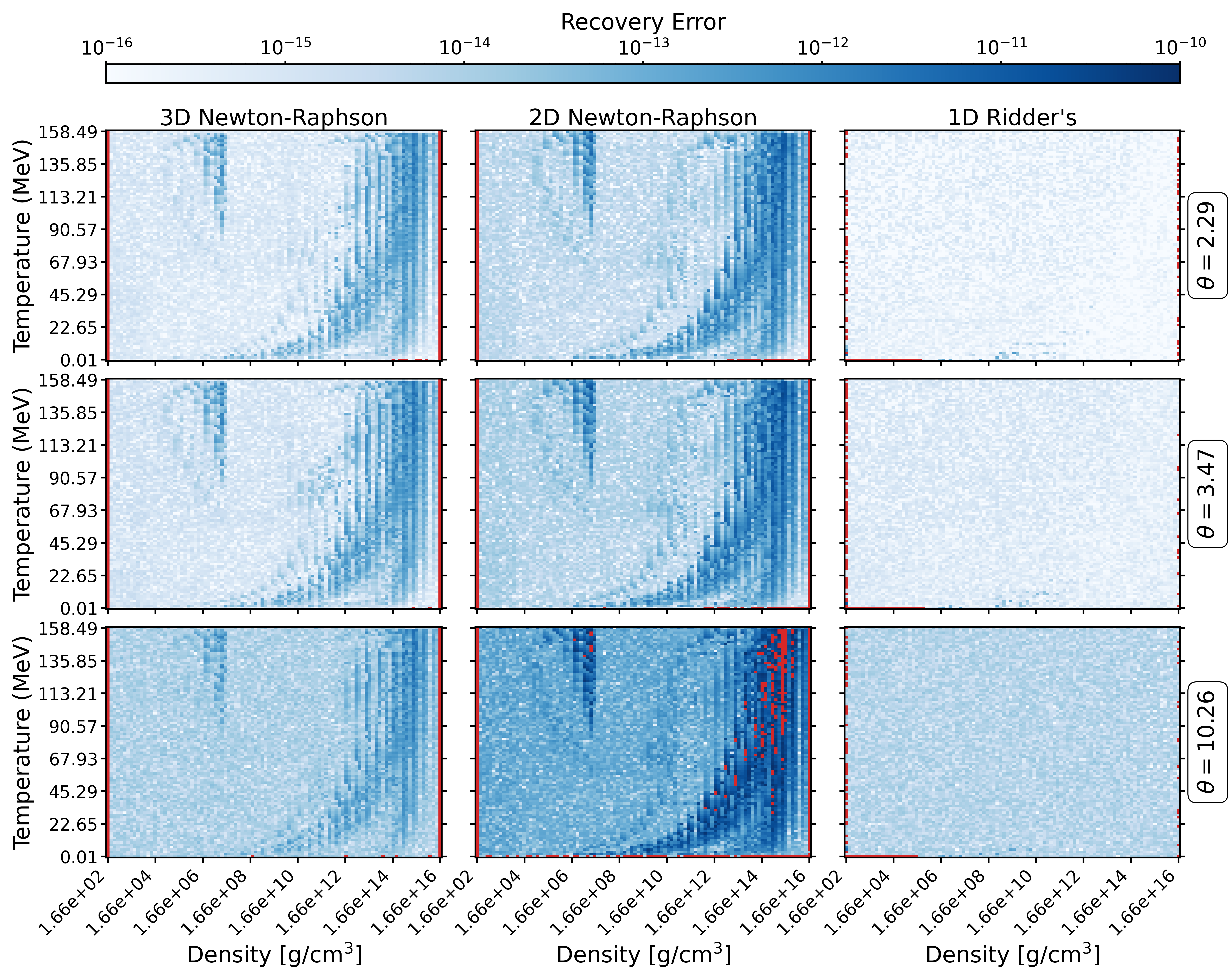}
\caption{Accuracy of \texttt{con2prim} in the $(\rho, T)$ plane for different schemes for the \texttt{DD2} EOS. The regions colored red are those where \texttt{con2prim} failed i.e. \texttt{con2prim} did not converge or converged with a recovery error greater than $10^{-10}$. The regions colored blue are those where \texttt{con2prim} converged with an accuracy of at least $10^{-10}$, and the shade of blue represents the recovery error with the darker shades representing higher recovery errors. The top panels show the results for different schemes for {\em Case~A}~: flat spacetime with velocity $v=(0.9, 0, 0)$ and generalized Lorentz factor $\Theta \simeq 2.29$. The middle panels show the results for {\em Case~B}~: Cartesian Kerr-Schild spacetime with $v=(0.2, 0, 0)$ and $\Theta \simeq 3.47$. The bottom panels show the results for {\em Case~C}~:  Cartesian Kerr-Schild spacetime with $v=(0.9, 0, 0)$ and $\Theta \simeq 10.26$. We find that 3D Newton-Raphson and 2D Newton-Raphson show non-uniform convergence behavior where the recovery error is higher ($\lesssim 10^{-10}$) for densities $\gtrsim 10^{14} \, \mathrm{g/cm}^3$ and lower ($\lesssim 10^{-12}$) elsewhere. On the other hand, the recovery error is uniformly low everywhere for 1D Ridders' method with values $\lesssim 10^{-15}$ at $\Theta=2.29$ and $\lesssim 10^{-13}$ at $\Theta=10.26$. Thus, 1D Ridders' method is more accurate, especially in high-density regions.}
\label{fig:convergence_all}
\end{figure*}

\begin{figure*}
\centering
\includegraphics[width=1.0\textwidth]{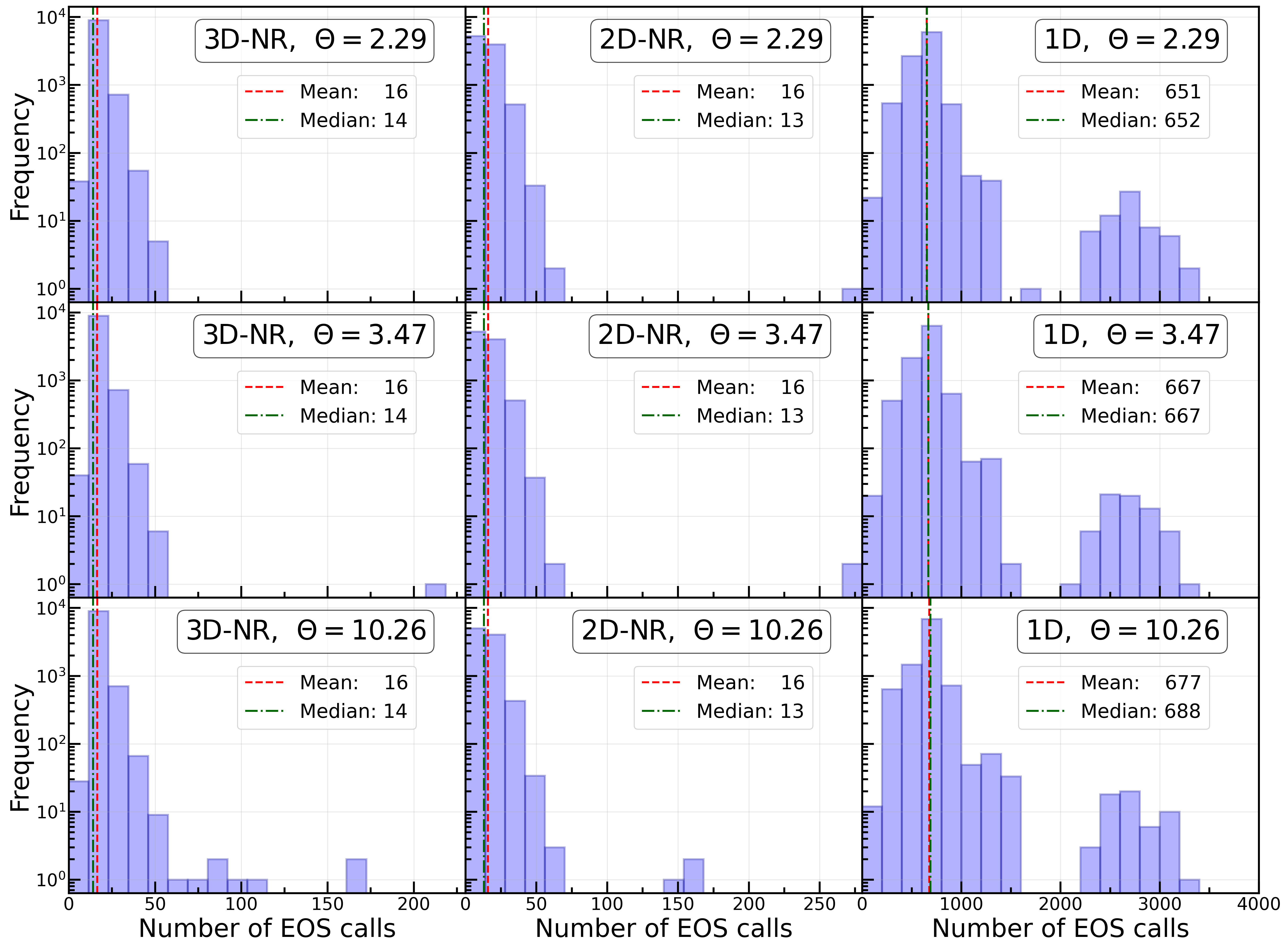}
\caption{Computational costs associated with different \texttt{con2prim} schemes in terms of number of EOS interpolations required. We plot the histograms of frequency vs number of EOS calls using 20 bins between 0 to 230 (280, 4000) for 3D Newton-Raphson (2D Newton-Raphson, 1D Ridders'). We also show the mean and the median number of EOS calls using dashed red and green lines respectively on top of the histograms. We compare the costs of 3D Newton-Raphson, 2D Newton-Raphson and 1D Ridders' method for $\Theta = 2.29$ in the top panels, $\Theta = 3.47$ in the middle panels and $\Theta = 10.26$ in the bottom panels. We find that 1D Ridders' method, even though the most accurate and robust, is also the most expensive among all schemes, with median EOS calls $\gtrsim 650$. On the other hand, 3D Newton-Raphson and 2D Newton-Raphson, though less robust, are much cheaper with median EOS calls $\lesssim 15$.}
\label{fig:cost_all}
\end{figure*}

We compare the different \texttt{con2prim} schemes in terms of correctness and computational cost/speed. We create a set of 100 $\rho$ values uniformly distributed in log-scale in the range $[\rho_\mathrm{min}, \rho_\mathrm{max}]$, as well as a set of 100 $T$ and 10 $Y_e$ values, uniformly distributed in linear-scale in the range $[T_\mathrm{min}, T_\mathrm{max}]$ and $[Y_{e,\mathrm{min}}, Y_{e,\mathrm{max}}]$ respectively. The subscripts $\mathrm{min}$ and $\mathrm{max}$ denote the minimum and maximum value of the corresponding quantity in the EOS table. Thus, we have a total of 
$10^5$ samples in the $(\rho, T, Y_e)$ space. We \textit{choose} a particular velocity $v^i$, and calculate the pressure $P$ and specific energy density $u$ using the EOS table. We thus have $10^5$ samples of primitives $(\rho, u, v^i)$.

For each primitive sample $\mathcal{P}_0 = (\rho, u, v^i)$, we calculate the corresponding conservatives $\mathcal{C}_0 = (\rho^*, S_i, \hat{e})$ using the straightforward \texttt{prim2con} operation, see Eqs.~(\ref{eq:p2c1}), (\ref{eq:p2c2}) and (\ref{eq:p2c3}). We then perturb the initial primitives  $\mathcal{P}_0$ by $10\%$ 
and then use them as the initial guess for the \texttt{con2prim} using various schemes\footnote{ To check the robustness, we also performed tests with perturbations of $20\%$ and $30\%$ for the initial primitives, and found small extra costs in the Newton-Raphson methods, but similar accuracy.}. We note that while the 3D and 2D Newton-Raphson schemes are dependent on the initial guess, the 1D Ridders' method does not need any initial guess. We perform \texttt{con2prim} with all three schemes using conservatives $\mathcal{C}_0$ as input, which give us the output primitives $\mathcal{P}_1$ with a tolerance of $10^{-10}$. We then perform another \texttt{prim2con} with primitives $\mathcal{P}_1$ as input, which gives us the output conservatives $\mathcal{C}_1$. We consider \texttt{con2prim} as successful if the following conditions are met:
\begin{enumerate}
    \item The \texttt{con2prim} $\mathcal{C}_0 \rightarrow \mathcal{P}_1$ converges with an accuracy of at least $10^{-10}$.
    This means that the recovery error $\varepsilon$ falls below $10^{-10}$ during the iterations. The recovery error is defined in Eq.~\ref{eq:error-3DNR} for 3D Newton-Raphson, Eq.~(\ref{eq:error-2DNR}) for 2D Newton-Raphson and Eq.~(\ref{eq:error-1DRidders}) for 1D Ridders' method.
    \item The difference between $\mathcal{C}_0$ and $\mathcal{C}_1$ is at most $10^{-10}$. We define the difference between $\mathcal{C}_0$ and $\mathcal{C}_1$ as follows:
    \begin{equation}
    \begin{aligned}
            ||\mathcal{C}_1 - \mathcal{C}_0|| &= 
            \frac{1}{5} \Bigg[ \left|\frac{\rho^*_1 - \rho^*_0}{\rho^*_0}\right|  
            + \left|\frac{\hat{e}_1 - \hat{e}_0}{\hat{e}_0}\right| \\
            &\qquad + \Sigma_i \left|\frac{S_{i,1} - S_{i,0}}{S_{i,0}}\right|
            \Bigg] 
    \end{aligned}
    \end{equation}
    where the factor 5 is due to an average over the 5 components of the conservative variables.
\end{enumerate}

In Fig.~\ref{fig:convergence_all}, we show the results for \texttt{con2prim} success/failure in the $(\rho, T)$ plane for $Y_e = 0.14$. We chose this value of $Y_e$ because it is one of the discrete tabulated $Y_e$ entries in our tests that is in the regime expected in a neutron star merger, but nothing in the following is in any way sensitive to this choice. Blue points in Fig.~\ref{fig:convergence_all}
 represent success while red points indicate failure, and higher intensity of blue represents higher recovery error. We find that the results 
are rather insensitive to the exact value of  $Y_e$. We perform the tests in both flat  and curved spacetime. For the curved spacetime tests, we use the Cartesian Kerr-Schild (CKS) metric, see e.g. \cite{wiltshire09}, with black-hole mass $M_\mathrm{BH} = 3 M_\odot$ and dimensionless spin parameter $\chi = 0.7$. We choose a total of 3 spacetime metrics as follows:
\begin{enumerate}
    \item {\em Case A}: Flat metric and velocity: ($v^x = 0.9, v^y = 0.0, v^z = 0.0$), which results in $\Theta = 2.2942$.
        In this case, the deviation of the Lorentz factor from unity is entirely due to the velocity.
    \item {\em Case B}: CKS metric at coordinates: ($x=10, y=0, z=0$) and velocity: ($v^x = 0.2, v^y = 0.0, v^z = 0.0$), which results in $\Theta = 3.4742$.  In this case, the Lorentz factor has a substantial contribution from the curved spacetime.
    \item {\em Case C}: CKS metric at coordinates: ($x=120, y=0, z=0$) and velocity: ($v^x = 0.9, v^y = 0.0, v^z = 0.0$), which results in $\Theta = 10.2606$. In this case, the Lorentz factor has substantial contributions from both the large velocity and the curved spacetime.
\end{enumerate}
We find that for the flat spacetime case with $\Theta \sim 2.29$ and the curved spacetime case with $\Theta \sim 3.47$, all 3 \texttt{con2prim} methods show an excellent success rate of $\gtrsim98\%$. Some failures happen at the edges of the table because the density/pressure falls out of the table bounds during the iterations.  In the curved spacetime case with $\Theta \sim 10.26$, both 1D and 3D methods show excellent success rate of  $\gtrsim98\%$ as before, 
however the 2D method shows slightly degraded convergence with a success rate of $\sim 95\%$. This degraded convergence for the 2D case happens for some high-density, high-temperature points that fail to converge in addition to the table edges.

In order to estimate the cost of each \texttt{con2prim} method, we use the number of EOS table interpolations required to reach convergence because the majority of cost incurred in tabulated EOS based \texttt{con2prim} methods comes from table interpolations. We refer to these table interpolations as the ``number of EOS calls" hereafter for simplicity. It should be noted that the number of EOS calls is in general $\geq 1$ for each iteration. For the calculations, we only include points where \texttt{con2prim} is successful. We show the frequency distribution of the number of EOS calls required for convergence of each \texttt{con2prim} method in Fig.~\ref{fig:cost_all}. We find that the 3D and the 2D Newton-Raphson methods require  only $\sim 15$ EOS calls to converge for different $\Theta$ cases on average. On the other hand, the 1D Ridders' method requires $\gtrsim 650$ EOS calls for convergence on average, which is $\sim40$ times higher than 3D/2D methods. The reason the 1D method is so costly is that it needs to invert the EOS table to find the temperature for a given specific internal energy \textit{at every iteration}. This is not the case with 2D/3D methods because temperature is an independent iteration variable and thus the inversion of the EOS table is not required.


We have also performed the accuracy and cost analysis for the 3 \texttt{con2prim} methods with a chosen tolerance of $10^{-8}$ and $10^{-12}$, in addition to $10^{-10}$ as discussed above. For brevity, we do not show the figures for these cases and only report the results for {\em Case C} with $\Theta \approx 10.26$ which is the most extreme case. We also only report the results for 3D Newton-Raphson and 1D Ridders' since these are the methods we use for production simulations. We find that the success rate of 3D Newton-Raphson is $\sim98\%$ for both $10^{-8}$ and $10^{-10}$ tolerance, while it decreases to $\sim97\%$ for the $10^{-12}$ tolerance. This decrease happens due to additional failures in high density, high temperature regions. On the other hand, Ridders' method shows an excellent success rate of $\gtrsim99\%$ in all cases. The median number of EOS calls remains 14 for 3D Newton-Raphson for both the $10^{-8}$ and $10^{-10}$ tolerance, while it increases slightly to 16 for the $10^{-12}$ case. However, the median number of EOS calls for Ridders' method is 638, 688 and 715 for the tolerance of $10^{-8}$, $10^{-10}$ and $10^{-12}$, respectively. Therefore, 3D Newton-Raphson is cheap irrespective of the tolerance chosen, but shows a higher failure rate at $10^{-12}$. Meanwhile, Ridders' method is essentially fail-safe irrespective of the tolerance chosen, but it becomes progressively expensive with decreasing tolerance.

The results we have discussed so far are based on a given set of initial conditions that we have \textit{chosen} to test. However, in an actual neutron star merger simulation, a variety of conditions can be encountered which vary depending on the phase of the simulation. It is therefore important to test these \texttt{con2prim} methods for robustness and cost in an actual production setup. For the production simulations, we choose to use the 3D Newton-Raphson as the \textit{primary} \texttt{con2prim} method because it is very cheap and also rather robust. However, the robustness of this method depends on the initial guess for primitives (which we take as the primitives from the last timestep). Thus, it is possible that this method fails when quantities change rapidly, especially during the merger and post-merger phase. In such cases, we use the 1D Ridders' method as the \textit{backup} \texttt{con2prim} method 
because it is independent of the initial guess and thus very robust even when there are rapid changes in physical quantities. Thus, we have a less robust yet cheap primary method, and an expensive yet very robust backup method. This is similar to the approach used in Eulerian (magneto)-hydrodynamics codes~\citep{Noble_2006, Siegel_2018, kastaun21, Shankar_2023, Cupp_2026}.  We do not use the 2D Newton-Raphson method in production simulations because, even though it has similar cost as the 3D Newton-Raphson method, it is less robust at higher Lorentz factors.

\section{Binary neutron star merger simulation}\label{sec:BNS_simulation}

\begin{figure*}
\centering
\includegraphics[width=1.0\textwidth]{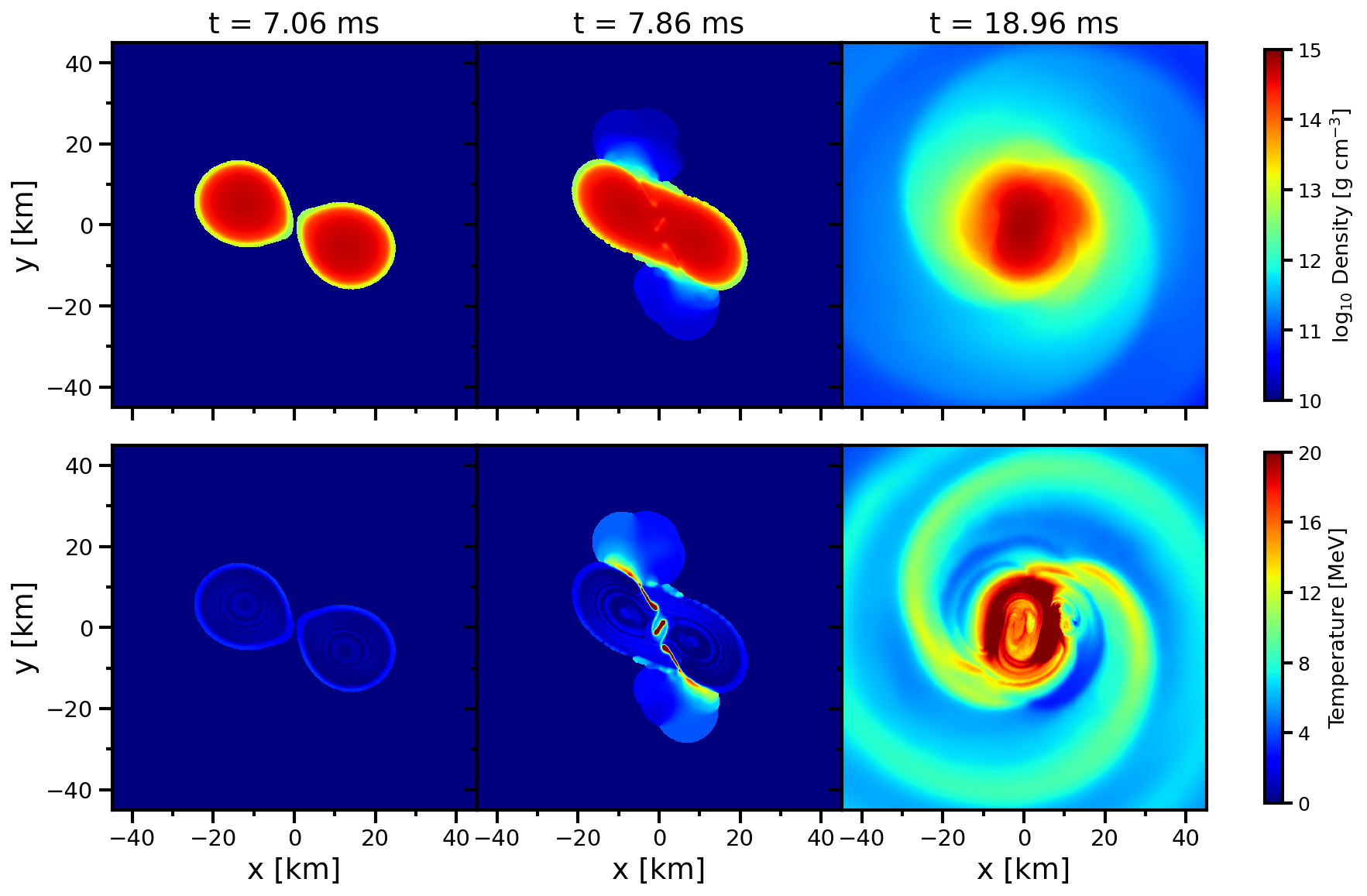}
\caption{2D cross-sections of density and temperature of the binary neutron star merger simulation just before the merger at $7.06 \, \mathrm{ms}$, just after the merger at $7.86 \, \mathrm{ms}$ and the final simulation time at $18.96 \, \mathrm{ms}$. The cross-sections are in the $xy$-plane at $z=0$ for the simulation with 3 million particles. We show the density in the top panels and the temperature (capped at 20 MeV for visibility) in the bottom panels.}
\label{fig:simulation_rho_T}
\end{figure*}

We simulate a $2\times1.3 M_{\odot}$, irrotational binary neutron star system using the DD2 EOS \citep{hempel10}.  We generate the hydro and spacetime initial data for this system using the \texttt{FUKA} library \citep{papenfort21} with \texttt{GRHayl}~\citep{Cupp_2026} support for stellarcollapse.org format EOS tables. To represent the cold inspiral neutron stars as best as we can, we set a constant initial temperature within the neutron stars using the minimum available temperature in the EOS table, T=0.01 MeV, and the beta-equilibrium electron fraction. After data generation with \texttt{FUKA}, we translate this to particle initial data using the Artificial Pressure Method (APM), originally suggested in a Newtonian context \citep{rosswog20a}, but adapted to the general relativistic case more recently \citep{rosswog21a,diener22a,rosswog23a}. The initial coordinate separation of the neutron stars is $45 \, \mathrm{km}$. We evolve the hydrodynamics using 1, 2 and 3 million particles as mentioned earlier, and the spacetime with the BSSN formulation \citep{baumgarte10,rezzolla13a,shibata16} of the Einstein equations with additional constraint damping terms \citep{etienne24,biswas26} using a Cartesian grid with fixed mesh refinement. Initially the Cartesian grid consists of 7 refinement levels with a fine resolution of $0.325 \ (0.369, \ 0.477) \, \mathrm{km}$ for the 3 (2, 1) million particle simulation, and a coarse resolution of $20.8 \ (23.6, \ 29.8) \, \mathrm{km}$. After merger, when the central density increases, another refinement level is added in all cases. This is done so that the spacetime resolution can match the increased SPH resolution. The boundaries are at $\pm 2268 \, \mathrm{km}$ in each direction for all simulations. We use the 3rd order Runge-Kutta method for time integration with a CFL factor of 0.2 for the simulations. 

The initial electron fraction in the neutron stars is determined by the (neutrino-less) beta-equilibrium condition. Unlike Eulerian methods, we do not need to actively advect $Y_e$, instead each fluid particle simply carries its own $Y_e$ along with it as it moves. Since we do not include weak interactions in these simulations, 
each particle has an associated equilibrium $Y_e$ at the start of the simulation, which remains constant throughout the evolution. The spatial distribution of $Y_e$ is thus purely governed by the motion of the particles in the current simulation. 

The central density of the neutron stars is $\sim 5.55 \times 10^{14} \, \mathrm{g/cm}^3$ initially, and saturates to $\sim 7 \times 10^{14} \, \mathrm{g/cm}^3$ at the end of the simulation time for all simulations. The merger time, which we calculate as the time at which the gravitational wave luminosity peaks, varies depending on the resolution, with the merger times being $5.26 \, \mathrm{ms}$, $6.84 \, \mathrm{ms}$ and $7.79 \, \mathrm{ms}$ for simulations with 1, 2 and 3 million particles respectively. The remnant does not collapse to a black hole by the end of the simulation, with the minimum lapse saturating close to $0.52$. The final ejecta mass is $6 \times 10^{-3} M_{\odot}$ for the simulation with 3 million particles, and it varies slightly for simulations with 2 million and 1 million particles with ejecta masses of $9 \times 10^{-3} M_{\odot}$ and $7 \times 10^{-3} M_{\odot}$ respectively. 

At the employed resolutions, the simulations have not yet converged as can be seen in merger times and ejecta masses. With the mixed methodology (adaptive grid, particles, particle$\leftrightarrow$grid mapping etc.), it is difficult to theoretically predict the order at which the method should converge. SPH is, however, expected to converge at second order~\citep[e.g.][]{rosswog2026} and this is also the order it has been found to converge for the Hamiltonian constraint~\citep{diener22a}.

We show the rest mass density and temperature for the simulation with 3 million particles in Fig.~\ref{fig:simulation_rho_T}. We show snapshots just before the merger at $t\sim7.1 \, \mathrm{ms}$, slightly after the merger at $t \sim 7.9 \, \mathrm{ms}$ and the final time at $t \sim 19 \, \mathrm{ms}$.  During the inspiral phase, the temperature in the neutron stars rises from the initial cold and uniform value of $0.01 \, \mathrm{MeV}$, and oscillates close to $1 \, \mathrm{MeV}$ at the center and close to $3 \, \mathrm{MeV}$ at the surface. The maximum temperature attained during the inspiral phase remains $\lesssim 3.5 \, \mathrm{MeV}$ until the neutron stars make contact. This is substantially lower than in corresponding simulations with Eulerian codes, see e.g. \cite{hammond21,gittins25}.
This is closely related to the Lagrangian nature of SPH and its  ability to deal with vacuum. Since the basic (computing frame) density estimate, Eq.~(\ref{eq:rhostar}), involves a summation over positive masses and positive definite kernels, the density at a particle position is always guaranteed to be
positive. Therefore, it is always safe to divide by densities as in the momentum evolution equation, Eq.~(\ref{eq:GR_momentum_evolution}), and the energy evolution equation, Eq.~(\ref{eq:GR_energy_evolution}). The SPH particles then adjust at the stellar surface to a numerical (i.e. resolution-dependent)
equilibrium configuration which transitions to vacuum (i.e. the absence of particles) naturally. No different treatment for ``fluid" and ``vacuum"/``atmosphere" is needed.
Overall, a carefully prepared initial condition
of a neutron star remains nearly perfectly in hydrostatic equilibrium, see, for example,
Fig.~10 in \cite{rosswog21a}.
 While the stellar surfaces do not need any special treatment,
they are still a numerically challenging region since the density drops, on a very short length scale, by orders of magnitude. Therefore, the particles in the surface continuously
have to adjust to the changing pull from the companion star and this motion causes also in SPH's case a temperature increase in the stellar surface, however, at a substantially lower extent than in Eulerian methods. \\ 
During the merger, the temperature in local vortices in the shear layer can exceed $50 \, \mathrm{MeV}$. During further evolution, the temperature also rises sharply at the outskirts of the remnant when the material flung out from the shear layer makes contact with the outer layers of the remnant. Spiral arms with temperatures $\sim 10 \, \mathrm{MeV}$ are formed around $9 \, \mathrm{ms}$ outside the remnant which evolve in an asymmetric fashion until $13 \, \mathrm{ms}$ and become generally symmetric thereafter. The temperatures in the spiral arms remain close to $\sim 10 \, \mathrm{MeV}$ at the outskirts and rise to $\sim 15 \, \mathrm{MeV}$ closer to the remnant, while the maximum temperature in the remnant itself rises to $\sim 40 \, \mathrm{MeV}$ towards the end of the simulation.

\begin{figure}
\centering
\includegraphics[width=1.0\columnwidth]{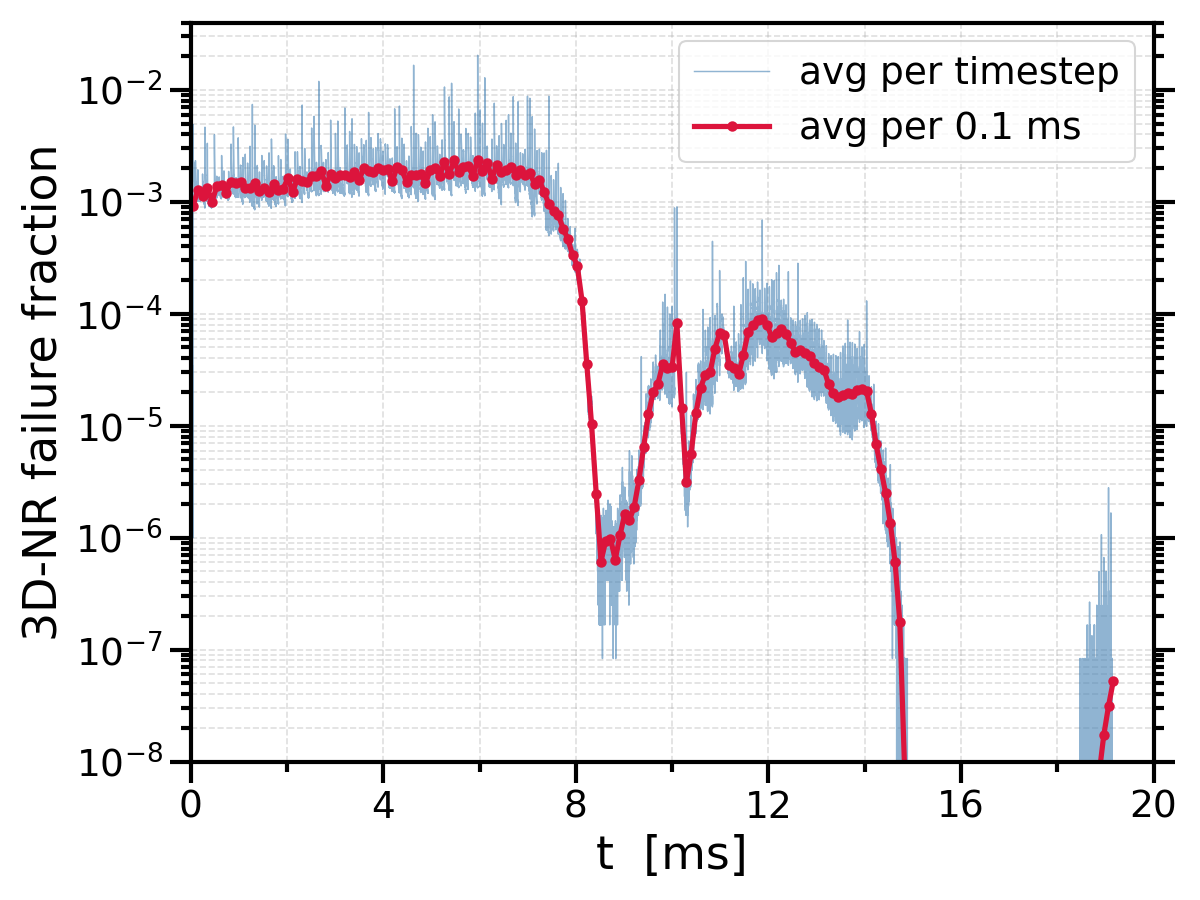}
\caption{Failure fraction of our primary 3D Newton-Raphson based \texttt{con2prim} method as a function of time. At every sub-step, we define the failure fraction as the ratio of number of particles where 3D Newton-Raphson fails and backup 1D Ridders' method is needed to the total number of particles. The blue line shows the average failure fraction at every timestep. The red line shows the average failure fraction over every $0.1 \, \mathrm{ms}$ of time. We find that the average failure rate of 3D Newton-Raphson is $\sim0.1\%$ in the pre-merger phase and $\lesssim0.01\%$ in the post-merger phase.}
\label{fig:ridders_ratio}
\end{figure}

As mentioned before, we use the method based on 3D Newton-Raphson as the primary \texttt{con2prim} and the method based on 1D Ridders' as the backup \texttt{con2prim} in the presented simulations. We find that the time required for 3D Newton-Raphson is typically only $\sim3\%$ of the total hydrodynamics time during the inspiral phase, whereas for this phase the 1D Ridders' method requires $\sim30\%$ of the total hydrodynamics time. We find that the failure rate of 3D Newton-Raphson is typically $\sim 0.1-1\%$ in the inspiral phase, $<0.1\%$ during the early post-merger phase and $\sim 0.0\%$ in the late post-merger phase. This reduction of 3D Newton-Raphson failure rate is likely related to the disappearance of the sharp neutron star surfaces in the post-merger phase.
We show the failure rate as a function of time in Fig.~\ref{fig:ridders_ratio}. In case both 3D Newton-Raphson and subsequently 1D Ridders' methods fail, we perform the \texttt{con2prim} with Ridders' method again with a tolerance of $10^{-8}$ instead of $10^{-10}$. This last resort fallback is typically triggered in $\lesssim10$ out of a million particles in the inspiral phase, and occasionally ($\sim1$ out of a million particles) in the post-merger phase. In all the instances that 3D Newton-Raphson failed, we find that the 1D Ridders' method (with the last resort fallback) was successful in the recovery of primitives. Thus, the 1D Ridders' method is very robust in the production environment in all the phases of the simulation. Additionally, the amount of time taken by 1D Ridders' method is similar for different phases of the simulation because it is independent of the initial primitives. Additionally, despite being substantially more expensive than the 3D Newton-Raphson, the 1D Ridders' method is still relatively cheap and typically takes $\sim8\%$ of the \textit{total} compute time. This is because the hydrodynamics uses approximately only one-third of the total compute time, whereas another one-third is spent on spacetime evolution and the remaining one-third is used for other operations such as mapping, prolongation, restriction, and input-output.   

We have also performed simulations with 1 million particles with \texttt{con2prim} tolerance of $10^{-8}$ and $10^{-12}$, which is in addition to the simulation with tolerance $10^{-10}$. We find that in going from tolerance $10^{-8}$ to $10^{-10}$, the \texttt{con2prim} time increases by $\sim20\%$ and overall runtime increases by only $\sim1\%$. However, going from tolerance $10^{-10}$ to $10^{-12}$, the \texttt{con2prim} time increases by $\sim375\%$ and overall runtime increases by $\sim10\%$. This happens because for simulations with tolerance $10^{-8}$ and $10^{-10}$, the Ridders' fallback is triggered at max $\lesssim0.1-1\%$ of the time. On the other hand, for the simulation with tolerance $10^{-12}$, the Ridders' fallback is triggered $\lesssim10\%$ of the time, especially in the inspiral phase. The simulations with tolerance $10^{-8}$, $10^{-10}$ and $10^{-12}$ all show almost identical results in the inspiral phase. At the post-merger phase, the simulations with tolerance $10^{-8}$ and $10^{-12}$ show differences of $\lesssim3\%$ and $\lesssim1\%$ respectively compared to the simulation with tolerance $10^{-10}$, when comparing the maximum densities at particle positions. Thus, the differences are slightly larger and the total simulation cost does not increase much up to a tolerance of $10^{-10}$. On the other hand, for a lower tolerance of $10^{-12}$, the differences are small, yet the cost increases by $\sim10\%$. This justifies our choice of tolerance of $10^{-10}$ for the simulations. These results are consistent with the conclusions from the accuracy and cost analysis in Sec.~\ref{sec:results}.

We have a residual check at convergence in the Ridders' method, but there is no such residual check in the 3D Newton-Raphson method due of a lack of appropriate scaling factors in this case. In order to ensure that the results are not affected by this and keeping future simulations in mind, we have also performed additional simulations where we use a hybrid approach: if 3D Newton-Raphson converges to a solution with pressure $P_{\text{NR}}$, we \textit{always} perform an additional Ridders' \texttt{con2prim} with the pressure guess $P_{\text{NR}}$ and the initial bracket  [$0.1P_{\text{NR}}, 10P_{\text{NR}}$]. If this hybrid 3D Newton-Raphson is not successful, we use the original initial pressure bracket from the table bounds. This removes any uncertainty regarding the residuals. We find that the results with the hybrid approach are almost identical in the inspiral phase, and show small differences ($\lesssim 1\%$) in the post-merger phase, for example when comparing the maximum densities at particle positions.

\section{Conclusions And discussion}\label{sec:conclusions}
Binary neutron star merger simulations 
should involve, apart from accurate and robust numerical methods, a faithful representation of the involved physics to allow for meaningful comparisons with and interpretations of observations. The equation of state in particular adds important microphysics to the simulations and it has effects on essentially all of the potential observables including gravitational waves and electromagnetic counterparts, see e.g. \cite{baiotti17,radice18a,shibata19,baiotti19,sarin21}. While some equations of state are available in analytic form, the most realistic equations of state based on microphysical calculations usually come in the form of pre-calculated tables. The required interpolation, in such tables, makes the conservative-to-primitive (\texttt{con2prim}) transformation, that is usually required in numerical relativity codes, very challenging.

In this work, we present three  \texttt{con2prim} transformation schemes for tabulated equations of state that we have developed for the  equations of the general relativistic Lagrangian hydrodynamics code \texttt{SPHINCS\_BSSN}. This development  is necessary because \texttt{SPHINCS\_BSSN} uses a different set of conservative variables and transformation equations compared to Eulerian (magneto-)hydrodynamics codes, and hence the existing \texttt{con2prim} algorithms from the Eulerian codes cannot be applied. In particular, we describe a 3D Newton-Raphson method, a 2D Newton-Raphson method and a 1D scheme based on Ridders' method for the recovery of primitive variables. We use the generalized Lorentz factor $\Theta$, the specific enthalpy $\mathcal{E}$ and the temperature $T$ as root-finding variables for the 3D Newton-Raphson scheme. For the 2D Newton-Raphson case, we use $\Theta$ and $T$ as root-finding variables, and, for the 1D Ridders' method, we use the pressure $P$.  

We test the robustness and computational cost of the three schemes using $10^5$ samples in the $(\rho, T, Y_e)$ space for the DD2 EOS for both flat and curved spacetimes, corresponding to various generalized Lorentz factors. We find that all the methods show an excellent \texttt{con2prim} success rate of $\gtrsim 98\%$ for $\Theta \lesssim 3.5$. However, for high generalized Lorentz factors $\Theta \gtrsim 10$, the 2D Newton-Raphson method shows a slightly degraded success rate of $\sim 95\%$, whereas 3D Newton-Raphson and 1D Ridders' methods still maintain the excellent success rate of $\sim 98\%$. We also find that while 3D Newton-Raphson shows a success rate of $\sim98\%$ in both the parameter study as well as the BNS simulation, 1D Ridders' method shows a success rate of $\sim99\%$ in the parameter study, but $\sim100\%$ in the BNS simulation: it never failed in the BNS simulation we performed in this work, except for the $\sim10$ particles in a million where the last resort fallback with Ridders' with a tolerance of $10^{-8}$ was successful. The reason for this could be either (1) the parameter space in the BNS simulation seldom touches the points of failure that we found in the parameter study, or (2) in the parameter study, we check for convergence as well as perform another \texttt{prim2con} to check consistency with initial conservatives, while in the BNS simulation we only check for convergence but do not perform another \texttt{prim2con} in order to reduce computational costs. In terms of computational cost, we find that the 3D and 2D Newton-Raphson methods require, on average, only $\sim 15$ EOS calls to converge, while the 1D Ridders' method requires, on average, $\gtrsim 650$ EOS calls to converge which is $\sim 40$ times more expensive. Despite being more expensive, the 1D Ridders' method is more robust because it does not depend on an initial guess. This is not the case with the Newton-Raphson method which depends on an initial guess and can fail if the initial guess is too far from the correct solution or the derivatives are not sufficiently accurate.

To demonstrate that the presented \texttt{con2prim} methods are robust enough for practical simulations, we perform binary neutron star merger simulations where we use the faster 3D Newton-Raphson method as the primary \texttt{con2prim} method, and the expensive, yet robust, 1D Ridders' method as a ``parachute" \texttt{con2prim} method. As a last resort fallback, we use Ridders' method with a tolerance of $10^{-8}$ instead of $10^{-10}$ which has always worked.  We do not use the 2D Newton-Raphson method in the production simulations, because it is less reliable at higher generalized Lorentz factors, and has no obvious advantage compared to the other two methods. We perform the simulations with 1, 2 and 3 million SPH particles using the DD2 EOS. These are the first
\texttt{SPHINCS\_BSSN} neutron star merger simulations with a tabulated finite-temperature equation of state. 
We find that, throughout the full simulation, the 3D Newton-Raphson method is successful $\gtrsim 99\%$ of the time, see Fig.~\ref{fig:ridders_ratio}. For the $\lesssim1\%$ of the time that it fails, the 1D Ridders' method is always able to successfully recover the primitives.
The combination of primary 3D Newton-Raphson and backup 1D Ridders' method for \texttt{con2prim} is computationally inexpensive and typically requires only $\lesssim 1\%$ of the total evolution time. On the other hand, simulations run only with 1D Ridders' method require $\sim 8\%$ of the total evolution time for \texttt{con2prim}. The success of these simulations demonstrates the robustness of our primary-backup strategy in all phases of the simulation. This is not unexpected since similar approaches have been successfully used in magnetohydrodynamics simulations which have additional complexities in the recovery of primitive variables due to the presence of magnetic fields~\citep{Siegel_2018}. \\
While it is usually considered as the ``gold standard" for a numerical relativity simulation, the use of tabulated EOSs also comes with  challenges. First, the density and temperature in the ejecta can become so low that they fall outside the table bounds, requiring special treatment such as density/temperature floors or the use of additional EOS in these regions. If such regions are a major concern, one may alternatively turn to analytical fits of the cold parts of the EOS, but enhanced with physics-based, analytical thermal contributions to pressure and internal energy \citep{raithel19,raithel21a,most21,biswas26}. Second, the \texttt{con2prim} recovery can become challenging in non-smooth
parts of a table, e.g. where  phase transitions occur.
 Tabulated EOSs are also more expensive than an analytic EOS. None of these challenges, however, is specific to \texttt{SPHINCS\_BSSN} or to Lagrangian hydrodynamics, these are difficulties for numerical relativity simulations in general. The ability to use tabulated EOSs
 from now on opens up a wide range of possibilities for future work with \texttt{SPHINCS\_BSSN}, including the exploration of high-density matter physics (with minimal code changes) or the inclusion of neutrino physics.

\begin{acknowledgments}

Swapnil Shankar and Stephan Rosswog (SR) acknowledge the support 
by the European Research Council (ERC) Advanced Grant INSPIRATION under the European Union’s Horizon 2020 research and innovation programme (Grant agreement No. 101053985) and by Deutsche Forschungsgemeinschaft (DFG, German Research Foundation) under Germany's Excellence Strategy – EXC~2121 ``Quantum Universe'' – 390833306. SR further benefited from the
support of the Knut and Alice Wallenberg Foundation 
under grant Dnr.~KAW~2019.0112 and by the Swedish Research Council (VR) under grant number 2020-05044, by the research environment grant “Gravitational Radiation and Electromagnetic Astrophysical Transients” (GREAT) funded by the Swedish Research Council (VR) under Dnr 2016-06012. \\
The calculations were performed in part at the NHR Center NHR@ZIB, jointly supported
by the Federal Ministry of Education and Research and the state
governments participating in the NHR (www.nhr-verein.de/unsere-
partner), at the SUNRISE HPC facility supported by the Technical
Division at the Department of Physics, Stockholm University, and
on the HUMMEL2 cluster funded by the Deutsche Forschungsgemeinschaft (498394658). Special thanks go to Mikica Kocic (SU),
Thomas Orgis and Hinnerk Stüben (both UHH) for their excellent
support. Portions of this research were conducted with high performance computational resources provided by the Louisiana Optical Network Infrastructure (http://www.loni.org).

\end{acknowledgments}


\newpage
\appendix

\section{Derivatives for the Jacobian matrix in 3D Newton-Raphson}\label{appendix:jacobian_3DNR}
Here we provide explicitly the derivatives that are needed for the
Jacobian, Eq.~(\ref{eq:Jacobian_3DNR}). We use the standard trilinear interpolation to interpolate the pressure, $P_\mathrm{EOS}$, and the specific internal energy, $u_\mathrm{EOS}$, from the EOS table at a given $(\rho, T, Y_e)$. In order to calculate the derivatives such as $\left(\frac{\partial P_{\rm EOS}}{\partial \rho} \right)$ numerically from the EOS table, we use the first order backward finite differences. In the subsequent calculations shown here, we use it to  calculate the derivatives $\frac{\partial P_{\rm EOS}}{\partial \rho}, \frac{\partial P_{\rm EOS}}{\partial T}, \frac{\partial u_{\rm EOS}}{\partial \rho}$ and $\frac{\partial u_{\rm EOS}}{\partial T}$ numerically from the EOS table.
\begin{eqnarray}
\frac{\partial f^1}{\partial \Theta}
&=& \frac{\partial}{\partial \Theta}\left[ \Theta \mathcal{E} \rho^* \hat{e}  - \rho^* S^2 - \mathcal{E}^2 \rho^* + \Theta \mathcal{E} \sqrt{-g}\, P_\text{EOS} - \frac{g^{0j}S_j}{g^{00}} \left( \Theta \mathcal{E} \rho^* - \rho^* g^{0j} S_j \right) \right] \nonumber \\
&=& \mathcal{E} \rho^* \hat{e} + \mathcal{E} \sqrt{-g} P_{\rm EOS} + \Theta \mathcal{E} \sqrt{-g} \frac{\partial P_{\rm EOS}}{\partial\Theta} - \mathcal{E} \rho^* \frac{g^{0j} S_j}{g^{00}} \nonumber \\
&=& \mathcal{E} \rho^* \hat{e} + \mathcal{E} \sqrt{-g} P_{\rm EOS} + \Theta \mathcal{E} \sqrt{-g} \frac{\partial P_{\rm EOS}}{\partial \rho} \frac{\partial \rho}{\partial \Theta} - \mathcal{E} \rho^* \frac{g^{0j} S_j}{g^{00}} \nonumber \\
&=& \mathcal{E} \rho^* \hat{e} + \mathcal{E} \sqrt{-g} P_{\rm EOS} + \mathcal{E} \sqrt{-g} \frac{\partial P_{\rm EOS}}{\partial \rho} \left[\frac{-\rho^*}{\sqrt{-g}\Theta^2}\right] - \mathcal{E} \rho^* \frac{g^{0j} S_j}{g^{00}},   \quad\,\, (\rho = \frac{\rho^*}{\Theta\sqrt{-g}})  \nonumber \\
&=& \mathcal{E} \rho^* \hat{e} + \mathcal{E} \sqrt{-g} P_{\rm EOS} -  \frac{\mathcal{E} \rho^*}{\Theta} \left(\frac{\partial P_{\rm EOS}}{\partial \rho} \right) - \mathcal{E} \rho^* \frac{g^{0j} S_j}{g^{00}}
\end{eqnarray}

%
We further have
\begin{eqnarray}
\frac{\partial f^1}{\partial \mathcal{E}} 
&=& \frac{\partial}{\partial \mathcal{E}}\left[ \Theta \mathcal{E} \rho^* \hat{e}  - \rho^* S^2 - \mathcal{E}^2 \rho^* + \Theta \mathcal{E} \sqrt{-g}\, P_\text{EOS} - \frac{g^{0j}S_j}{g^{00}} \left( \Theta \mathcal{E} \rho^* - \rho^* g^{0j} S_j \right) \right] \nonumber \\
&=& \Theta \rho^* \hat{e} - 2 \mathcal{E} \rho^* + \Theta \sqrt{-g} P_{\rm EOS} - \Theta \rho^* \frac{g^{0j} S_j}{g^{00}}
\end{eqnarray}
%
and
\begin{eqnarray}
\frac{\partial f^1}{\partial T}
&=& \frac{\partial}{\partial T}\left[ \Theta \mathcal{E} \rho^* \hat{e}  - \rho^* S^2 - \mathcal{E}^2 \rho^* + \Theta \mathcal{E} \sqrt{-g}\, P_\text{EOS} - \frac{g^{0j}S_j}{g^{00}} \left( \Theta \mathcal{E} \rho^* - \rho^* g^{0j} S_j \right) \right] 
= \Theta \mathcal{E} \sqrt{-g} \left(\frac{\partial P_{\rm EOS}}{\partial T}\right).
\end{eqnarray}
\begin{eqnarray}
\frac{\partial f^2}{\partial \Theta}
&=& \frac{\partial}{\partial \Theta} \left[\rho^* \Theta \mathcal{E} - \rho^* g^{0j} S_j + g^{00}P_{\text{EOS}} (\rho,T,Y_e)\sqrt{-g} + \rho^* g^{00} \hat{e} \right] 
= \rho^* \mathcal{E} + g^{00}\sqrt{-g} \frac{\partial P_{\rm EOS}}{\partial \rho} \frac{\partial \rho}{\partial \Theta}  \nonumber \\
&=& \rho^* \mathcal{E} + g^{00}\sqrt{-g} \frac{\partial P_{\rm EOS}}{\partial \rho} \left[\frac{-\rho^*}{\sqrt{-g}\Theta^2}\right] 
= \rho^* \mathcal{E} -  \frac{g^{00} \rho^*}{\Theta^2} \left(\frac{\partial P_{\rm EOS}}{\partial \rho} \right),  
\end{eqnarray}
where we have used Eq.(\ref{eq:rho_of_rhostar}). The remaining $f^2$ derivatives are
\begin{eqnarray}
\frac{\partial f^2}{\partial \mathcal{E}} 
&=& \frac{\partial}{\partial \mathcal{E}} \left[\rho^* \Theta \mathcal{E} - \rho^* g^{0j} S_j + g^{00}P_{\text{EOS}} (\rho,T,Y_e)\sqrt{-g} + \rho^* g^{00} \hat{e} \right]
= \rho^* \Theta
\end{eqnarray}
and
\begin{eqnarray}
\frac{\partial f^2}{\partial T}
&=& \frac{\partial}{\partial T} \left[\rho^* \Theta \mathcal{E} - \rho^* g^{0j} S_j + g^{00}P_{\text{EOS}} (\rho,T,Y_e)\sqrt{-g} + \rho^* g^{00} \hat{e} \right]
= g^{00} \sqrt{-g}\left(\frac{\partial P_{\rm EOS}}{\partial T} \right).
\end{eqnarray}
%

\begin{eqnarray}
\frac{\partial f^3}{\partial \Theta}
&=& \frac{\partial}{\partial \Theta}
     \left[ u - u_{\text{EOS}}(\rho,T,Y_e) \right]
= \frac{\partial u}{\partial \Theta}
   - \frac{\partial u_{\text{EOS}}(\rho,T,Y_e)}{\partial \Theta}
\end{eqnarray}
Recall that
\begin{eqnarray}
u &=& \mathcal{E} - 1 - \frac{P}{\rho} = \mathcal{E} - 1 - \frac{P_{\text{EOS}} \Theta \sqrt{-g}}{\rho^*} \nonumber,
\end{eqnarray}
therefore
\begin{eqnarray}
\frac{\partial u}{\partial \Theta}
&=& - \frac{\partial}{\partial \Theta}
    \left( \frac{\sqrt{-g}}{\rho^*} P_{\text{EOS}} \Theta \right) 
= - \frac{\sqrt{-g}}{\rho^*}
   \left( P_{\text{EOS}} + \Theta \frac{\partial P_{\text{EOS}}}{\partial \Theta} \right) 
= - \frac{\sqrt{-g}}{\rho^*}
   \left( P_{\text{EOS}} + \Theta \frac{\partial P_{\text{EOS}}}{\partial \rho} \frac{\partial \rho}{\partial\Theta} \right) \nonumber \\ 
&=& - \frac{\sqrt{-g}}{\rho^*}
   \left( P_{\text{EOS}} + \Theta \frac{\partial P_{\text{EOS}}}{\partial \rho} \left[ \frac{-\rho^*}{\sqrt{-g} \Theta^2} \right] \right)
= - \frac{\sqrt{-g} P_{\text{EOS}} }{\rho^*} + \frac{1}{\Theta} \left( \frac{\partial P_{\text{EOS}}}{\partial \rho} \right)
\end{eqnarray}
and
\begin{eqnarray}
\frac{\partial u_{\text{EOS}}(\rho,T,Y_e)}{\partial \Theta}
&=& \frac{\partial u_{\text{EOS}}}{\partial \rho}  \frac{\partial \rho}{\partial \Theta} 
= \frac{\partial u_{\text{EOS}}}{\partial \rho}  \left[ \frac{-\rho^*}{\sqrt{-g} \Theta^2} \right]
= - \frac{1}{\Theta^2} \frac{\rho^*}{\sqrt{-g}} \left( \frac{\partial u_{\text{EOS}}}{\partial \rho} \right),
\end{eqnarray}
%
\begin{eqnarray}
\frac{\partial f^3}{\partial \mathcal{E}}
&=& \frac{\partial}{\partial \mathcal{E}}
     \left[ u - u_{\text{EOS}}(\rho,T,Y_e) \right] 
= \frac{\partial}{\partial \mathcal{E}} \left[ \mathcal{E} - 1 - \frac{P_{\text{EOS}} \Theta \sqrt{-g}}{\rho^*} \right] = 1
\end{eqnarray}
and
\begin{eqnarray}
\frac{\partial}{\partial T}(f^3)
&=& \frac{\partial}{\partial T} \left[ u - u_{\text{EOS}}(\rho,T,Y_e) \right]
= \frac{\partial u}{\partial T} - \frac{\partial u_{\text{EOS}}}{\partial T} \nonumber.
\end{eqnarray}
Now
\begin{eqnarray}
\frac{\partial u}{\partial T}
&=& \frac{\partial}{\partial T} \left[ \mathcal{E} - 1 - \frac{P_{\text{EOS}} \Theta \sqrt{-g}}{\rho^*} \right] = \frac{\Theta \sqrt{-g}}{\rho^*} \frac{\partial P_{\text{EOS}}}{\partial T} \nonumber,
\end{eqnarray}
therefore
\begin{eqnarray}
\frac{\partial f^3}{\partial T}
&=& - \frac{\Theta \sqrt{-g}}{\rho^*} \left( \frac{\partial P_{\text{EOS}}}{\partial T} \right)
   - \left( \frac{\partial u_{\text{EOS}}}{\partial T} \right).
\end{eqnarray}

\section{Derivatives for the Jacobian matrix in 2D Newton-Raphson}\label{appendix:jacobian_2DNR}
Here we provide explicitly the derivatives that are needed for the
Jacobian, Eq.~(\ref{eq:Jacobian_2DNR}).
\begin{eqnarray}
\frac{\partial}{\partial \Theta}(f^1)
&=& \frac{\partial}{\partial \Theta} \left[ \Theta^2 ( B^2 + A) + g^{00} B^2 \right]  = 2\Theta \left( B^2 + A \right) + 2\Theta^2 B \frac{\partial B}{\partial \Theta} + 2 g^{00} B \frac{\partial B}{\partial \Theta} \nonumber \\
&=& 2 \Theta \left( B^2 + A\right) + 2 B \frac{\partial B}{\partial \Theta} \left( \Theta^2 + g^{00} \right)
\end{eqnarray}
Recall that 
\begin{equation}
    B = B(P) = g^{0j}S_j - g^{00}\left( \frac{\sqrt{-g}}{\rho^*} P + \hat{e} \right) \nonumber,
\end{equation}
therefore
\begin{eqnarray}
   \frac{\partial B}{\partial \Theta} 
   &=&  \frac{\partial B}{\partial P} \frac{\partial P}{\partial \Theta} = \frac{\partial B}{\partial P} \frac{\partial P}{\partial \rho} \frac{\partial \rho}{\partial \Theta}
   = \frac{\partial B}{\partial P} \left(\frac{\partial P_\mathrm{EOS}}{\partial \rho} \right) \left[ \frac{-\rho^*}{\sqrt{-g} \Theta^2} \right] \nonumber \\
   &=& \left[ \frac{-g^{00} \sqrt{-g}}{\rho^*} \right] \left(\frac{\partial P_\mathrm{EOS}}{\partial \rho} \right) \left[ \frac{-\rho^*}{\sqrt{-g} \Theta^2} \right]   
   = \frac{g^{00}}{\Theta^2} \left(\frac{\partial P_\mathrm{EOS}}{\partial \rho} \right), 
\end{eqnarray}
where we have used
\begin{equation}
\frac{\partial B}{\partial P} = \frac{-g^{00}\sqrt{-g}}{\rho^*}. 
\end{equation}
\begin{eqnarray}
\frac{\partial f^1}{\partial T}
&=& \frac{\partial}{\partial T}
     \left[ \Theta^2 ( B^2 + A) + g^{00} B^2 \right] 
= \Theta^2 \left[ 2 B \frac{\partial B}{\partial T} \right] + 2 g^{00} B \frac{\partial B}{\partial T}
= 2 B \frac{\partial B}{\partial T} \left( \Theta^2 + g^{00} \right),
\end{eqnarray}
where $\frac{\partial B}{\partial T}$ can be calculated as follows
\begin{eqnarray}
    \frac{\partial B}{\partial T} 
    &=& \frac{\partial B}{\partial P} \frac{\partial P_\mathrm{EOS}}{\partial T}  
    = \frac{-g^{00} \sqrt{-g}}{\rho^*} \left( \frac{\partial P_\mathrm{EOS}}{\partial T} \right).
\end{eqnarray}
%
For the $f^2$ derivatives we find
\begin{eqnarray}
\frac{\partial f^2}{\partial \Theta}
&=& \frac{\partial}{\partial \Theta}
     \left[ u - u_{\text{EOS}}(\rho,T,Y_e) \right] = \frac{\partial u}{\partial \Theta}
   - \frac{\partial u_{\text{EOS}}(\rho,T,Y_e)}{\partial \Theta}.
\end{eqnarray}
The specific energy $u$ can be expressed in terms of $\Theta$ and $P$ as follows (see Eq.~(49) in \cite{rosswog21a}):
\begin{equation}
    u = \frac{1}{\Theta} ( g^{0j}S_j - g^{00}\hat{e}) - \frac{\sqrt{-g}P_{\text{EOS}}}{\Theta \rho^*}(g^{00} + \Theta^2) - 1  \nonumber,
\end{equation}
therefore,
\begin{equation}
\frac{\partial u}{\partial \Theta}
= -\frac{1}{\Theta^2} \left[ g^{0j} S_j - g^{00} \hat{e}\right] 
+ \frac{\sqrt{-g} g^{00}}{\rho^*} \frac{P_\mathrm{EOS}}{\Theta^2}
- \frac{\sqrt{-g} g^{00}}{\rho^* \Theta} \left( \frac{\partial P_\mathrm{EOS}}{\partial \Theta} \right) 
- \frac{\sqrt{-g}}{\rho^*} P_\mathrm{EOS} 
- \frac{\sqrt{-g}}{\rho^*} \Theta \left( \frac{\partial P_\mathrm{EOS}}{\partial \Theta} \right) \nonumber.
\end{equation}
Substituting $\frac{\partial P_\mathrm{EOS}}{\partial\Theta} = \frac{-\rho^*}{\sqrt{-g}\Theta^2} \left( \frac{\partial P_\mathrm{EOS}}{\partial \rho}\right)$ and simplifying, we finally get
\begin{equation}
    \frac{\partial u}{\partial\Theta} = 
    -\frac{1}{\Theta^2} \left[ g^{0j} S_j - g^{00} \hat{e}\right] 
    + \frac{\sqrt{-g} g^{00}}{\rho^*} \frac{P_\mathrm{EOS}}{\Theta^2} 
    + \frac{g^{00}}{\Theta^3} \left( \frac{\partial P_\mathrm{EOS}}{\partial \rho} \right)
    - \frac{\sqrt{-g}}{\rho^*} P_\mathrm{EOS}
    + \frac{1}{\Theta} \left( \frac{\partial P_\mathrm{EOS}}{\partial \rho} \right)
\end{equation}
and
\begin{eqnarray}
\frac{\partial u_{\text{EOS}}(\rho,T,Y_e)}{\partial \Theta}
&=& \frac{\partial u_{\text{EOS}}}{\partial \rho}  \frac{\partial \rho}{\partial \Theta} = \frac{\partial u_{\text{EOS}}}{\partial \rho}  \left[ \frac{-\rho^*}{\sqrt{-g} \Theta^2} \right]  
= - \frac{1}{\Theta^2} \frac{\rho^*}{\sqrt{-g}} \left( \frac{\partial u_{\text{EOS}}}{\partial \rho} \right).
\end{eqnarray}
%
For $\partial f^2/\partial T$ we find
\begin{eqnarray}
\frac{\partial f^2}{\partial T}
&=& \frac{\partial}{\partial T} \left[ u - u_{\text{EOS}}(\rho,T,Y_e) \right] 
= \frac{\partial u}{\partial T} - \frac{\partial u_{\text{EOS}}}{\partial T} \nonumber
\end{eqnarray}
and
\begin{eqnarray}
\frac{\partial u}{\partial T}
&=& \frac{\partial}{\partial T} \left[ \frac{1}{\Theta} ( g^{0j}S_j - g^{00}\hat{e}) - \frac{\sqrt{-g}P_{\text{EOS}}}{\Theta \rho^*}(g^{00} + \Theta^2) - 1 \right] 
=  - \frac{\sqrt{-g}}{\Theta \rho^*} \left( g^{00} + \Theta^2 \right) \left( \frac{\partial P_{\text{EOS}}}{\partial T} \right)  \nonumber \\
&=& - \frac{\sqrt{-g}}{\Theta \rho^*} \left( g^{00} + \Theta^2 \right) \left( \frac{\partial P_{\text{EOS}}}{\partial T} \right) \nonumber.
\end{eqnarray}
Therefore
\begin{eqnarray}
\frac{\partial f^2}{\partial T}
&=& - \frac{\sqrt{-g}}{\Theta \rho^*} \left( g^{00} + \Theta^2 \right) \left( \frac{\partial P_{\text{EOS}}}{\partial T} \right)
   - \left( \frac{\partial u_{\text{EOS}}}{\partial T} \right).
\end{eqnarray}


\bibliography{astro_SKR_after_11_2025.bib}{}

@Article{tanvir17,
  author        = {{Tanvir}, N.~R. and {Levan}, A.~J. and {Gonz{a}lez-Fern{a}ndez}, C. and {Korobkin}, O. and {Mandel}, I. and {Rosswog}, S. and {Hjorth}, J. and {D'Avanzo}, P. and more},
  title         = {{The Emergence of a Lanthanide-rich Kilonova Following the Merger of Two Neutron Stars}},
  doi           = {10.3847/2041-8213/aa90b6},
  eid           = {L27},
  eprint        = {1710.05455},
  pages         = {L27},
  volume        = {848},
  adsurl        = {https://ui.adsabs.harvard.edu/abs/2017ApJ...848L..27T},
  archiveprefix = {arXiv},
  journal       = {\apjl},
  month         = oct,
  primaryclass  = {astro-ph.HE},
  year          = {2017},
}

@Article{baym18,
  author        = {{Baym}, Gordon and {Hatsuda}, Tetsuo and {Kojo}, Toru and {Powell}, Philip D. and {Song}, Yifan and {Takatsuka}, Tatsuyuki},
  title         = {{From hadrons to quarks in neutron stars: a review}},
  doi           = {10.1088/1361-6633/aaae14},
  eid           = {056902},
  eprint        = {1707.04966},
  number        = {5},
  pages         = {056902},
  volume        = {81},
  adsurl        = {https://ui.adsabs.harvard.edu/abs/2018RPPh...81e6902B},
  archiveprefix = {arXiv},
  journal       = {Rep. Prog. Phys.},
  month         = may,
  primaryclass  = {astro-ph.HE},
  year          = {2018},
}

@Article{price06,
  author  = {D.J. Price and S. Rosswog},
  title   = {Producing ultra-strong magnetic fields in neutron star mergers},
  pages   = {719},
  volume  = {312},
  journal = {Science},
  year    = {2006},
}

@Article{perego19a,
       author = {{Perego}, Albino and {Bernuzzi}, Sebastiano and {Radice}, David},
        title = "{Thermodynamics conditions of matter in neutron star mergers}",
      journal = {European Physical Journal A},
         year = 2019,
        month = aug,
       volume = {55},
       number = {8},
          eid = {124},
        pages = {124},
          doi = {10.1140/epja/i2019-12810-7},
archivePrefix = {arXiv},
       eprint = {1903.07898},
 primaryClass = {gr-qc},
       adsurl = {https://ui.adsabs.harvard.edu/abs/2019EPJA...55..124P}
}

@Article{radice18a,
  author        = {{Radice}, David and {Perego}, Albino and {Hotokezaka}, Kenta and {Fromm}, Steven A. and {Bernuzzi}, Sebastiano and {Roberts}, Luke F.},
  title         = {{Binary Neutron Star Mergers: Mass Ejection, Electromagnetic Counterparts, and Nucleosynthesis}},
  doi           = {10.3847/1538-4357/aaf054},
  eid           = {130},
  eprint        = {1809.11161},
  number        = {2},
  pages         = {130},
  volume        = {869},
  adsurl        = {https://ui.adsabs.harvard.edu/abs/2018ApJ...869..130R},
  archiveprefix = {arXiv},
  journal       = {\apj},
  month         = {Dec},
  primaryclass  = {astro-ph.HE},
  year          = {2018},
}

@Article{metzger20,
  author        = {{Metzger}, Brian D.},
  title         = {{Kilonovae}},
  doi           = {10.1007/s41114-019-0024-0},
  eid           = {1},
  eprint        = {1910.01617},
  pages         = {1},
  volume        = {23},
  adsurl        = {https://ui.adsabs.harvard.edu/abs/2019LRR....23....1M},
  archiveprefix = {arXiv},
  journal       = {Living Rev. Relativ.},
  primaryclass  = {astro-ph.HE},
  year          = {2020},
}

@Article{foucart23,
  author        = {{Foucart}, Francois},
  title         = {{Neutrino transport in general relativistic neutron star merger simulations}},
  doi           = {10.1007/s41115-023-00016-y},
  eid           = {1},
  eprint        = {2209.02538},
  pages         = {1},
  volume        = {9},
  adsurl        = {https://ui.adsabs.harvard.edu/abs/2023LRCA....9....1F},
  archiveprefix = {arXiv},
  journal       = {Living Rev. Comput. Astrophys.},
  primaryclass  = {astro-ph.HE},
  year          = {2023},
}

@Article{yang24,
  author        = {{Yang}, Yu-Han and {Troja}, Eleonora and {O'Connor}, Brendan and {Fryer}, Chris L. and {Im}, Myungshin and {Durbak}, Joe and more},
  title         = {{A lanthanide-rich kilonova in the aftermath of a long gamma-ray burst}},
  doi           = {10.1038/s41586-023-06979-5},
  eprint        = {2308.00638},
  number        = {8000},
  pages         = {742-745},
  volume        = {626},
  adsurl        = {https://ui.adsabs.harvard.edu/abs/2024Natur.626..742Y},
  archiveprefix = {arXiv},
  journal       = {Nature},
  month         = feb,
  primaryclass  = {astro-ph.HE},
  year          = {2024},
}

@Article{baryakhtar22,
  author        = {{Baryakhtar}, Masha and {Caputo}, Regina and {Croon}, Djuna and {Perez}, Kerstin and {Berti}, Emanuele and {Bramante}, Joseph and {Buschmann}, Malte and {Brito}, Richard and {Chen}, Thomas Y. and {Cole}, Philippa S. and {Coogan}, Adam and {East}, William E. and {Foster}, Joshua W. and {Galanis}, Marios and {Giannotti}, Maurizio and {Kavanagh}, Bradley J. and {Laha}, Ranjan and {Leane}, Rebecca K. and {Lehmann}, Benjamin V. and {Marques-Tavares}, Gustavo and {McDonald}, Jamie and {Ng}, Ken K.~Y. and {Raj}, Nirmal and {Sagunski}, Laura and {Sakstein}, Jeremy and {Sathyaprakash}, B.~S. and {Shandera}, Sarah and {Siemonsen}, Nils and {Simon}, Olivier and {Sinha}, Kuver and {Singh}, Divya and {Singh}, Rajeev and {Sun}, Chen and {Sun}, Ling and {Takhistov}, Volodymyr and {Tsai}, Yu-Dai and {Vitagliano}, Edoardo and {Vitale}, Salvatore and {Yang}, Huan and {Zhang}, Jun},
  title         = {{Dark Matter In Extreme Astrophysical Environments}},
  doi           = {10.48550/arXiv.2203.07984},
  eid           = {arXiv:2203.07984},
  eprint        = {2203.07984},
  pages         = {arXiv:2203.07984},
  adsurl        = {https://ui.adsabs.harvard.edu/abs/2022arXiv220307984B},
  archiveprefix = {arXiv},
  journal       = {arXiv e-prints},
  month         = mar,
  primaryclass  = {hep-ph},
  year          = {2022},
}

@Article{laguna93a,
  author  = {{Laguna}, P. and {Miller}, W.~A. and {Zurek}, W.~H.},
  title   = {{Smoothed particle hydrodynamics near a black hole}},
  doi     = {10.1086/172321},
  pages   = {678-685},
  volume  = {404},
  adsurl  = {http://adsabs.harvard.edu/abs/1993ApJ...404..678L},
  journal = {\apj},
  month   = feb,
  year    = {1993},
}

@Misc{compose,
  author = {{CompOSE}},
  title  = {{CompStar Online Supernova Equations of State}},
  url    = {https://compose.obspm.fr},
  year   = {2022},
}

@Article{diener22a,
  author        = {{Diener}, Peter and {Rosswog}, Stephan and {Torsello}, Francesco},
  title         = {{Simulating neutron star mergers with the Lagrangian Numerical Relativity code SPHINCS\_BSSN}},
  doi           = {10.1140/epja/s10050-022-00725-7},
  eid           = {74},
  eprint        = {2203.06478},
  number        = {4},
  pages         = {74},
  volume        = {58},
  adsurl        = {https://ui.adsabs.harvard.edu/abs/2022EPJA...58...74D},
  archiveprefix = {arXiv},
  journal       = {Eur. Phys. J. A},
  month         = apr,
  primaryclass  = {astro-ph.HE},
  year          = {2022},
}

@Book{bona09,
  author    = {Bona, Carles and Palenzuela-Luque, Carlos and Bona-Casas, Carles},
  title     = {{Elements of Numerical Relativity and Relativistic Hydrodynamics: From Einstein' s Equations to Astrophysical Simulations}},
  publisher = {Springer},
  address   = {Berlin, Heidelberg},
  doi       = {10.1007/978-3-642-01164-1},
  year      = {2009},
}

@Article{zrake13,
  author        = {{Zrake}, J. and {MacFadyen}, A.~I.},
  title         = {{Magnetic Energy Production by Turbulence in Binary Neutron Star Mergers}},
  doi           = {10.1088/2041-8205/769/2/L29},
  eid           = {L29},
  eprint        = {1303.1450},
  pages         = {L29},
  volume        = {769},
  adsurl        = {http://adsabs.harvard.edu/abs/2013ApJ...769L..29Z},
  archiveprefix = {arXiv},
  journal       = {\apjl},
  month         = jun,
  primaryclass  = {astro-ph.HE},
  year          = {2013},
}

@Article{rosswog09b,
  author  = {S. Rosswog},
  title   = {Astrophysical Smooth Particle Hydrodynamics},
  pages   = {78-104},
  volume  = {53},
  journal = {\nar},
  year    = {2009},
}

@Article{rastinejad22,
  author        = {{Rastinejad}, Jillian C. and {Gompertz}, Benjamin P. and {Levan}, Andrew J. and {Fong}, Wen-fai and {Nicholl}, Matt and {Lamb}, Gavin P. and {Malesani}, Daniele B. and {Nugent}, Anya E. and more},
  title         = {{A kilonova following a long-duration gamma-ray burst at 350 Mpc}},
  doi           = {10.1038/s41586-022-05390-w},
  eprint        = {2204.10864},
  number        = {7939},
  pages         = {223-227},
  volume        = {612},
  adsurl        = {https://ui.adsabs.harvard.edu/abs/2022Natur.612..223R},
  archiveprefix = {arXiv},
  journal       = {Nature},
  month         = dec,
  primaryclass  = {astro-ph.HE},
  year          = {2022},
}

@Article{price12a,
  author        = {{Price}, D.~J.},
  title         = {{Smoothed particle hydrodynamics and magnetohydrodynamics}},
  doi           = {10.1016/j.jcp.2010.12.011},
  eprint        = {1012.1885},
  pages         = {759-794},
  volume        = {231},
  adsurl        = {http://adsabs.harvard.edu/abs/2012JCoPh.231..759P},
  archiveprefix = {arXiv},
  journal       = {J. Comput. Phys.},
  month         = feb,
  primaryclass  = {astro-ph.IM},
  year          = {2012},
}

@Article{arcavi17,
  author        = {{Arcavi}, Iair and {Hosseinzadeh}, Griffin and {Howell}, D. Andrew and {McCully}, Curtis and {Poznanski}, Dovi and {Kasen}, Daniel and {Barnes}, Jennifer and {Zaltzman}, Michael and {Vasylyev}, Sergiy and {Maoz}, Dan and {Valenti}, Stefano},
  title         = {{Optical emission from a kilonova following a gravitational-wave-detected neutron-star merger}},
  doi           = {10.1038/nature24291},
  eprint        = {1710.05843},
  number        = {7678},
  pages         = {64-66},
  volume        = {551},
  adsurl        = {https://ui.adsabs.harvard.edu/abs/2017Natur.551...64A},
  archiveprefix = {arXiv},
  journal       = {\nat},
  month         = nov,
  primaryclass  = {astro-ph.HE},
  year          = {2017},
}

@Article{fortin21,
  author        = {{Fortin}, Jean-Fran{\c{c}}ois and {Guo}, Huai-Ke and {Harris}, Steven P. and {Kim}, Doojin and {Sinha}, Kuver and {Sun}, Chen},
  title         = {{Axions: From magnetars and neutron star mergers to beam dumps and BECs}},
  doi           = {10.1142/S0218271821300020},
  eid           = {2130002},
  eprint        = {2102.12503},
  number        = {7},
  pages         = {2130002},
  volume        = {30},
  adsurl        = {https://ui.adsabs.harvard.edu/abs/2021IJMPD..3030002F},
  archiveprefix = {arXiv},
  journal       = {Int. J. Mod. Phys. D},
  month         = oct,
  primaryclass  = {hep-ph},
  year          = {2021},
}

@Article{emma22,
  author        = {{Emma}, Mattia and {Schianchi}, Federico and {Pannarale}, Francesco and {Sagun}, Violetta and {Dietrich}, Tim},
  title         = {{Numerical Simulations of Dark Matter Admixed Neutron Star Binaries}},
  doi           = {10.3390/particles5030024},
  eprint        = {2206.10887},
  number        = {3},
  pages         = {273-286},
  volume        = {5},
  adsurl        = {https://ui.adsabs.harvard.edu/abs/2022Parti...5..273E},
  archiveprefix = {arXiv},
  journal       = {Particles},
  month         = jul,
  primaryclass  = {gr-qc},
  year          = {2022},
}

@Article{freiburghaus99b,
  author  = {C. Freiburghaus and S. Rosswog and F.-K. Thielemann},
  title   = {R-Process in Neutron Star Mergers},
  pages   = {L121},
  volume  = {525},
  journal = {\apj},
  year    = {1999},
}

@Article{das21,
  author        = {{Das}, H.~C. and {Kumar}, Ankit and {Patra}, S.~K.},
  title         = {{Dark matter admixed neutron star as a possible compact component in the GW190814 merger event}},
  doi           = {10.1103/PhysRevD.104.063028},
  eid           = {063028},
  eprint        = {2109.01853},
  number        = {6},
  pages         = {063028},
  volume        = {104},
  adsurl        = {https://ui.adsabs.harvard.edu/abs/2021PhRvD.104f3028D},
  archiveprefix = {arXiv},
  journal       = {\prd},
  month         = sep,
  primaryclass  = {astro-ph.HE},
  year          = {2021},
}

@Book{baumgarte10,
  author    = {Baumgarte, T. W. and Shapiro, S. L.},
  title     = {{Numerical Relativity: Solving {E}instein's Equations on the Computer}},
  publisher = {Cambridge University Press},
  address   = {Cambridge; New York},
  adsurl    = {http://adsabs.harvard.edu/abs/2010nure.book.....B},
  gbs_id    = {dxU1OEinvRUC},
  year      = {2010},
}

@Article{cowan21,
  author        = {{Cowan}, John J. and {Sneden}, Christopher and {Lawler}, James E. and {Aprahamian}, Ani and {Wiescher}, Michael and {Langanke}, Karlheinz and {Martinez-Pinedo}, Gabriel and {Thielemann}, Friedrich-Karl},
  title         = {{Origin of the heaviest elements: The rapid neutron-capture process}},
  doi           = {10.1103/RevModPhys.93.015002},
  eid           = {015002},
  eprint        = {1901.01410},
  number        = {1},
  pages         = {015002},
  volume        = {93},
  adsurl        = {https://ui.adsabs.harvard.edu/abs/2021RvMP...93a5002C},
  archiveprefix = {arXiv},
  journal       = {Rev. Mod. Phys.},
  month         = jan,
  primaryclass  = {astro-ph.HE},
  year          = {2021},
}

@Article{gittins25,
  author        = {{Gittins}, F. and {Matur}, R. and {Andersson}, N. and {Hawke}, I.},
  title         = {{Problematic systematics in neutron-star merger simulations}},
  doi           = {10.1103/PhysRevD.111.023049},
  eid           = {023049},
  eprint        = {2409.13468},
  number        = {2},
  pages         = {023049},
  volume        = {111},
  adsurl        = {https://ui.adsabs.harvard.edu/abs/2025PhRvD.111b3049G},
  archiveprefix = {arXiv},
  journal       = {\prd},
  month         = jan,
  primaryclass  = {gr-qc},
  year          = {2025},
}

@Article{rosswog03a,
  author  = {{Rosswog}, S. and {Liebend{\"o}rfer}, M.},
  title   = {{High-resolution calculations of merging neutron stars - II. Neutrino emission}},
  doi     = {10.1046/j.1365-8711.2003.06579.x},
  pages   = {673-689},
  volume  = {342},
  adsurl  = {http://adsabs.harvard.edu/abs/2003MNRAS.342..673R},
  journal = {\mnras},
  month   = jul,
  year    = {2003},
}

@Article{rosswog15c,
  author        = {{Rosswog}, S.},
  title         = {{SPH Methods in the Modelling of Compact Objects}},
  doi           = {10.1007/lrca-2015-1},
  eid           = {1},
  eprint        = {1406.4224},
  volume        = {1},
  archiveprefix = {arXiv},
  journal       = {Living Rev. Comput. Astrophys.},
  primaryclass  = {astro-ph.IM},
  year          = {2015},
}

@Article{symbalisty82,
  author   = {{Symbalisty}, E. and {Schramm}, D.~N.},
  title    = {{Neutron Star Collisions and the r-Process}},
  pages    = {143},
  volume   = {22},
  adsurl   = {https://ui.adsabs.harvard.edu/abs/1982ApL....22..143S},
  journal  = {Astrophys. Lett.},
  month    = jan,
  year     = {1982},
}

@Article{zhang07,
  author   = {{Zhang}, B. and {Zhang}, B.-B. and {Liang}, E.-W. and {Gehrels}, N. and {Burrows}, D.~N. and {Meszaros}, P.},
  title    = {{Making a Short Gamma-Ray Burst from a Long One: Implications for the Nature of GRB 060614}},
  doi      = {10.1086/511781},
  eprint   = {astro-ph/0612238},
  pages    = {L25-L28},
  volume   = {655},
  adsurl   = {http://adsabs.harvard.edu/abs/2007ApJ...655L..25Z},
  journal  = {\apjl},
  month    = jan,
  year     = {2007},
}

@Article{holz05,
  author        = {{Holz}, Daniel E. and {Hughes}, Scott A.},
  title         = {{Using Gravitational-Wave Standard Sirens}},
  doi           = {10.1086/431341},
  eprint        = {astro-ph/0504616},
  number        = {1},
  pages         = {15-22},
  volume        = {629},
  adsurl        = {https://ui.adsabs.harvard.edu/abs/2005ApJ...629...15H},
  archiveprefix = {arXiv},
  journal       = {\apj},
  month         = aug,
  primaryclass  = {astro-ph},
  year          = {2005},
}

@Book{alcubierre08,
  author    = {{Alcubierre}, M.},
  title     = {{Introduction to 3+1 Numerical Relativity}},
  publisher = {Oxford University Press},
  year      = {2008},
}

@Article{dessart09,
  author   = {{Dessart}, L. and {Ott}, C.~D. and {Burrows}, A. and {Rosswog}, S. and {Livne}, E.},
  title    = {{Neutrino Signatures and the Neutrino-Driven Wind in Binary Neutron Star Mergers}},
  doi      = {10.1088/0004-637X/690/2/1681},
  pages    = {1681-1705},
  volume   = {690},
  adsurl   = {http://adsabs.harvard.edu/abs/2009ApJ...690.1681D},
  journal  = {\apj},
  month    = jan,
  year     = {2009},
}

@Article{rezzolla11,
  author        = {{Rezzolla}, L. and {Giacomazzo}, B. and {Baiotti}, L. and {Granot}, J. and {Kouveliotou}, C. and {Aloy}, M.~A.},
  title         = {{The Missing Link: Merging Neutron Stars Naturally Produce Jet-like Structures and Can Power Short Gamma-ray Bursts}},
  doi           = {10.1088/2041-8205/732/1/L6},
  eid           = {L6},
  eprint        = {1101.4298},
  pages         = {L6},
  volume        = {732},
  adsurl        = {http://adsabs.harvard.edu/abs/2011ApJ...732L...6R},
  archiveprefix = {arXiv},
  journal       = {\apjl},
  month         = may,
  primaryclass  = {astro-ph.HE},
  year          = {2011},
}

@Article{eichler89,
  author  = {D. Eichler and M. Livio and T. Piran and D. N. Schramm},
  title   = {Nucleosynthesis, Neutrino Bursts and $\gamma$-Ray from Coalescing Neutron Stars},
  pages   = {126},
  volume  = {340},
  journal = {Nature},
  year    = {1989},
}

@ARTICLE{etienne24,
       author = {{Etienne}, Zachariah B.},
        title = "{Improved moving-puncture techniques for compact binary simulations}",
      journal = {\prd},
         year = 2024,
        month = sep,
       volume = {110},
       number = {6},
          eid = {064045},
        pages = {064045},
          doi = {10.1103/PhysRevD.110.064045},
archivePrefix = {arXiv},
       eprint = {2404.01137},
 primaryClass = {gr-qc},
       adsurl = {https://ui.adsabs.harvard.edu/abs/2024PhRvD.110f4045E}
}

@Article{rosswog18a,
  author        = {{Rosswog}, S. and {Sollerman}, J. and {Feindt}, U. and {Goobar}, A. and {Korobkin}, O. and {Wollaeger}, R. and {Fremling}, C. and {Kasliwal}, M.~M.},
  title         = {{The first direct double neutron star merger detection: Implications for cosmic nucleosynthesis}},
  doi           = {10.1051/0004-6361/201732117},
  eid           = {A132},
  eprint        = {1710.05445},
  pages         = {A132},
  volume        = {615},
  adsurl        = {http://adsabs.harvard.edu/abs/2018A%26A...615A.132R},
  archiveprefix = {arXiv},
  journal       = {\aap},
  month         = {jul},
  primaryclass  = {astro-ph.HE},
  year          = {2018},
}

@Article{aguilera24,
  author        = {{Aguilera-Miret}, Ricard and {Palenzuela}, Carlos and {Carrasco}, Federico and {Rosswog}, Stephan and {Vigan{\`o}}, Daniele},
  title         = {{Delayed jet launching in binary neutron star mergers with realistic initial magnetic fields}},
  doi           = {10.1103/PhysRevD.110.083014},
  eid           = {083014},
  eprint        = {2407.20335},
  number        = {8},
  pages         = {083014},
  volume        = {110},
  adsurl        = {https://ui.adsabs.harvard.edu/abs/2024PhRvD.110h3014A},
  archiveprefix = {arXiv},
  journal       = {\prd},
  month         = oct,
  primaryclass  = {astro-ph.HE},
  year          = {2024},
}

@Article{aguilera25,
       author = {{Aguilera-Miret}, Ricard and {Christian}, Jan-Erik and {Rosswog}, Stephan and {Palenzuela}, Carlos},
        title = "{Robustness of Magnetic Field Amplification in Neutron Star Mergers}",
      journal = {\mnras},
         year = 2025,
        month = aug,
       volume = {542},
       number = {4},
        pages = {3067-3077},
          doi = {10.1093/mnras/staf1291},
archivePrefix = {arXiv},
       eprint = {2504.10604},
 primaryClass = {astro-ph.HE},
       adsurl = {https://ui.adsabs.harvard.edu/abs/2025MNRAS.542.3067A}
}

@Article{aguilera22,
  author        = {{Aguilera-Miret}, Ricard and {Vigan{\`o}}, Daniele and {Palenzuela}, Carlos},
  title         = {{Universality of the Turbulent Magnetic Field in Hypermassive Neutron Stars Produced by Binary Mergers}},
  doi           = {10.3847/2041-8213/ac50a7},
  eid           = {L31},
  eprint        = {2112.08406},
  number        = {2},
  pages         = {L31},
  volume        = {926},
  adsurl        = {https://ui.adsabs.harvard.edu/abs/2022ApJ...926L..31A},
  archiveprefix = {arXiv},
  journal       = {\apjl},
  month         = feb,
  primaryclass  = {gr-qc},
  year          = {2022},
}

@Book{shibata16,
  author    = {{Shibata}, Masaru},
  title     = {{Numerical Relativity}},
  doi       = {10.1142/9692},
  publisher = {World Scientific},
  adsurl    = {https://ui.adsabs.harvard.edu/abs/2016nure.book.....S},
  year      = {2016},
}

@Article{rosswog07c,
  author  = {S. Rosswog and D.J. Price},
  title   = {{MAGMA}: a magnetohydrodynamics code for merger applications},
  pages   = {915 - 931},
  volume  = {379},
  journal = {\mnras},
  year    = {2007},
}

@Article{palenzuela22,
  author        = {{Palenzuela}, Carlos and {Aguilera-Miret}, Ricard and {Carrasco}, Federico and {Ciolfi}, Riccardo and {Kalinani}, Jay Vijay and {Kastaun}, Wolfgang and {Mi{\~n}ano}, Borja and {Vigan{\`o}}, Daniele},
  title         = {{Turbulent magnetic field amplification in binary neutron star mergers}},
  doi           = {10.1103/PhysRevD.106.023013},
  eid           = {023013},
  eprint        = {2112.08413},
  number        = {2},
  pages         = {023013},
  volume        = {106},
  adsurl        = {https://ui.adsabs.harvard.edu/abs/2022PhRvD.106b3013P},
  archiveprefix = {arXiv},
  journal       = {\prd},
  month         = jul,
  primaryclass  = {gr-qc},
  year          = {2022},
}

@Article{lattimer12a,
  author   = {{Lattimer}, J.~M.},
  title    = {{The Nuclear Equation of State and Neutron Star Masses}},
  doi      = {10.1146/annurev-nucl-102711-095018},
  pages    = {485-515},  
  volume   = {62},
  adsurl   = {https://ui.adsabs.harvard.edu/abs/2012ARNPS..62..485L},
  journal  = {Annual Review of Nuclear and Particle Science},
   month    = nov,
  year     = {2012},                                                            
}

@Article{lattimer16,
       author = {{Lattimer}, James M. and {Prakash}, Madappa},
        title = "{The equation of state of hot, dense matter and neutron stars}",
      journal = {\physrep},
         year = 2016,
        month = mar,
       volume = {621},
        pages = {127-164},
          doi = {10.1016/j.physrep.2015.12.005},
archivePrefix = {arXiv},
       eprint = {1512.07820},
 primaryClass = {astro-ph.SR},
       adsurl = {https://ui.adsabs.harvard.edu/abs/2016PhR...621..127L}
}

@ARTICLE{baiotti17,
       author = {{Baiotti}, Luca and {Rezzolla}, Luciano},
        title = "{Binary neutron star mergers: a review of Einstein{\textquoteright}s richest laboratory}",
      journal = {Reports on Progress in Physics},
         year = 2017,
        month = sep,
       volume = {80},
       number = {9},
          eid = {096901},
        pages = {096901},
          doi = {10.1088/1361-6633/aa67bb},
archivePrefix = {arXiv},
       eprint = {1607.03540},
 primaryClass = {gr-qc},
       adsurl = {https://ui.adsabs.harvard.edu/abs/2017RPPh...80i6901B}
}

@ARTICLE{baiotti19,
       author = {{Baiotti}, Luca},
        title = "{Gravitational waves from neutron star mergers and their relation to the nuclear equation of state}",
      journal = {Progress in Particle and Nuclear Physics},
         year = 2019,
        month = nov,
       volume = {109},
          eid = {103714},
        pages = {103714},
          doi = {10.1016/j.ppnp.2019.103714},
archivePrefix = {arXiv},
       eprint = {1907.08534},
 primaryClass = {astro-ph.HE},
       adsurl = {https://ui.adsabs.harvard.edu/abs/2019PrPNP.10903714B}
}

@ARTICLE{sarin21,
       author = {{Sarin}, Nikhil and {Lasky}, Paul D.},
        title = "{The evolution of binary neutron star post-merger remnants: a review}",
      journal = {General Relativity and Gravitation},
         year = 2021,
        month = jun,
       volume = {53},
       number = {6},
          eid = {59},
        pages = {59},
          doi = {10.1007/s10714-021-02831-1},
archivePrefix = {arXiv},
       eprint = {2012.08172},
 primaryClass = {astro-ph.HE},
       adsurl = {https://ui.adsabs.harvard.edu/abs/2021GReGr..53...59S}
}

@ARTICLE{most21,
       author = {{Most}, Elias R. and {Raithel}, Carolyn A.},
        title = "{Impact of the nuclear symmetry energy on the post-merger phase of a binary neutron star coalescence}",
      journal = {\prd},
         year = 2021,
        month = dec,
       volume = {104},
       number = {12},
          eid = {124012},
        pages = {124012},
          doi = {10.1103/PhysRevD.104.124012},
archivePrefix = {arXiv},
       eprint = {2107.06804},
 primaryClass = {astro-ph.HE},
       adsurl = {https://ui.adsabs.harvard.edu/abs/2021PhRvD.104l4012M}
}

@ARTICLE{shibata19,
       author = {{Shibata}, Masaru and {Hotokezaka}, Kenta},
        title = "{Merger and Mass Ejection of Neutron Star Binaries}",
      journal = {Annual Review of Nuclear and Particle Science},
         year = 2019,
        month = oct,
       volume = {69},
        pages = {41-64},
          doi = {10.1146/annurev-nucl-101918-023625},
archivePrefix = {arXiv},
       eprint = {1908.02350},
 primaryClass = {astro-ph.HE},
       adsurl = {https://ui.adsabs.harvard.edu/abs/2019ARNPS..69...41S}
}

@ARTICLE{jiang21,
       author = {{Jiang}, Nan and {Yagi}, Kent},
        title = "{Probing modified gravitational-wave propagation through tidal measurements of binary neutron star mergers}",
      journal = {\prd},
         year = 2021,
        month = jun,
       volume = {103},
       number = {12},
          eid = {124047},
        pages = {124047},
          doi = {10.1103/PhysRevD.103.124047},
archivePrefix = {arXiv},
       eprint = {2104.04442},
 primaryClass = {gr-qc},
       adsurl = {https://ui.adsabs.harvard.edu/abs/2021PhRvD.103l4047J}
}

@ARTICLE{sakstein17,
       author = {{Sakstein}, Jeremy and {Jain}, Bhuvnesh},
        title = "{Implications of the Neutron Star Merger GW170817 for Cosmological Scalar-Tensor Theories}",
      journal = {\prl},
         year = 2017,
        month = dec,
       volume = {119},
       number = {25},
          eid = {251303},
        pages = {251303},
          doi = {10.1103/PhysRevLett.119.251303},
archivePrefix = {arXiv},
       eprint = {1710.05893},
 primaryClass = {astro-ph.CO},
       adsurl = {https://ui.adsabs.harvard.edu/abs/2017PhRvL.119y1303S}
}

@BOOK{wiltshire09,
       author = {{Wiltshire}, David L. and {Visser}, Matt and {Scott}, Susan M.},
        title = "{The Kerr Spacetime}",
         year = 2009,
       adsurl = {https://ui.adsabs.harvard.edu/abs/2009kesp.book.....W}
}

@ARTICLE{hempel10,
       author = {{Hempel}, Matthias and {Schaffner-Bielich}, J{\"u}rgen},
        title = "{A statistical model for a complete supernova equation of state}",
      journal = {\nphysa},
         year = 2010,
        month = jun,
       volume = {837},
       number = {3-4},
        pages = {210-254},
          doi = {10.1016/j.nuclphysa.2010.02.010},
archivePrefix = {arXiv},
       eprint = {0911.4073},
 primaryClass = {nucl-th},
       adsurl = {https://ui.adsabs.harvard.edu/abs/2010NuPhA.837..210H}
}

@Article{levan24,
  author        = {{Levan}, Andrew J. and {Gompertz}, Benjamin P. and {Salafia}, Om Sharan and {Bulla}, Mattia and {Burns}, Eric and {Hotokezaka}, Kenta and more},
  title         = {{Heavy-element production in a compact object merger observed by JWST}},
  doi           = {10.1038/s41586-023-06759-1},
  eprint        = {2307.02098},
  number        = {8000},
  pages         = {737-741},
  volume        = {626},
  adsurl        = {https://ui.adsabs.harvard.edu/abs/2024Natur.626..737L},
  archiveprefix = {arXiv},
  journal       = {Nature},
  month         = feb,
  primaryclass  = {astro-ph.HE},
  year          = {2024},
}

@Article{oechslin02,
  author  = {{Oechslin}, R. and {Rosswog}, S. and {Thielemann}, F.-K.},
  title   = {{Conformally flat smoothed particle hydrodynamics application to neutron star mergers}},
  number  = {10},
  pages   = {103005},
  volume  = {65},
  adsurl  = {http://adsabs.harvard.edu/abs/2002PhRvD..65j3005O},
  journal = {\prd},
  month   = may,
  year    = {2002},
}

@Article{monaghan05,
  author  = {{Monaghan}, J.~J.},
  title   = {{Smoothed particle hydrodynamics}},
  pages   = {1703-1759},
  volume  = {68},
  journal = {Rep. Prog. Phys.},
  month   = aug,
  year    = {2005},
}

@Article{monaghan01,
  author  = {{Monaghan}, J.~J. and {Price}, D.~J.},
  title   = {{Variational principles for relativistic smoothed particle hydrodynamics}},
  doi     = {10.1046/j.1365-8711.2001.04742.x},
  pages   = {381-392},
  volume  = {328},
  adsurl  = {http://adsabs.harvard.edu/abs/2001MNRAS.328..381M},
  journal = {\mnras},
  month   = dec,
  year    = {2001},
}

@Article{raithel19,
  author        = {{Raithel}, Carolyn A. and {{\"O}zel}, Feryal and {Psaltis}, Dimitrios},
  title         = {{Finite-temperature Extension for Cold Neutron Star Equations of State}},
  doi           = {10.3847/1538-4357/ab08ea},
  eid           = {12},
  eprint        = {1902.10735},
  number        = {1},
  pages         = {12},
  volume        = {875},
  adsurl        = {https://ui.adsabs.harvard.edu/abs/2019ApJ...875...12R},
  archiveprefix = {arXiv},
  journal       = {\apj},
  month         = apr,
  primaryclass  = {astro-ph.HE},
  year          = {2019},
}

@ARTICLE{raithel21a,
       author = {{Raithel}, Carolyn A. and {Paschalidis}, Vasileios and {{\"O}zel}, Feryal},
        title = "{Realistic finite-temperature effects in neutron star merger simulations}",
      journal = {\prd},
         year = 2021,
        month = sep,
       volume = {104},
       number = {6},
          eid = {063016},
        pages = {063016},
          doi = {10.1103/PhysRevD.104.063016},
archivePrefix = {arXiv},
       eprint = {2104.07226},
 primaryClass = {astro-ph.HE},
       adsurl = {https://ui.adsabs.harvard.edu/abs/2021PhRvD.104f3016R}
}

@Article{springel10a,
  author        = {{Springel}, V.},
  title         = {{Smoothed Particle Hydrodynamics in Astrophysics}},
  doi           = {10.1146/annurev-astro-081309-130914},
  eprint        = {1109.2219},
  pages         = {391-430},
  volume        = {48},
  adsurl        = {http://adsabs.harvard.edu/abs/2010ARA\%26A..48..391S},
  archiveprefix = {arXiv},
  journal       = {\araa},
  month         = sep,
  primaryclass  = {astro-ph.CO},
  year          = {2010},
}

@Article{galeazzi13,
  author        = {{Galeazzi}, Filippo and {Kastaun}, Wolfgang and {Rezzolla}, Luciano and {Font}, Jose A.},
  title         = {{Implementation of a simplified approach to radiative transfer in general relativity}},
  doi           = {10.1103/PhysRevD.88.064009},
  eid           = {064009},
  eprint        = {1306.4953},
  number        = {6},
  pages         = {064009},
  volume        = {88},
  adsurl        = {https://ui.adsabs.harvard.edu/abs/2013PhRvD..88f4009G},
  archiveprefix = {arXiv},
  journal       = {\prd},
  month         = sep,
  primaryclass  = {gr-qc},
  year          = {2013},
}

@Article{papenfort21,
  author        = {{Papenfort}, L. Jens and {Tootle}, Samuel D. and {Grandcl{\'e}ment}, Philippe and {Most}, Elias R. and {Rezzolla}, Luciano},
  title         = {{New public code for initial data of unequal-mass, spinning compact-object binaries}},
  doi           = {10.1103/PhysRevD.104.024057},
  eid           = {024057},
  eprint        = {2103.09911},
  number        = {2},
  pages         = {024057},
  volume        = {104},
  adsurl        = {https://ui.adsabs.harvard.edu/abs/2021PhRvD.104b4057P},
  archiveprefix = {arXiv},
  journal       = {\prd},
  month         = jul,
  primaryclass  = {gr-qc},
  year          = {2021},
}

@ARTICLE{hammond21,
       author = {{Hammond}, P. and {Hawke}, I. and {Andersson}, N.},
        title = "{Thermal aspects of neutron star mergers}",
      journal = {\prd},
         year = 2021,
        month = nov,
       volume = {104},
       number = {10},
          eid = {103006},
        pages = {103006},
          doi = {10.1103/PhysRevD.104.103006},
archivePrefix = {arXiv},
       eprint = {2108.08649},
 primaryClass = {astro-ph.HE},
       adsurl = {https://ui.adsabs.harvard.edu/abs/2021PhRvD.104j3006H}
}

@Article{rosswog99,
  author  = {S. Rosswog and M. Liebend\"orfer and F.-K. Thielemann and M.B. Davies and W. Benz and T. Piran},
  title   = {Mass ejection in neutron star mergers},
  pages   = {499-526},
  volume  = {341},
  journal = {\aap},
  year    = {1999},
}

@Article{nissanke10,
  author        = {{Nissanke}, Samaya and {Holz}, Daniel E. and {Hughes}, Scott A. and {Dalal}, Neal and {Sievers}, Jonathan L.},
  title         = {{Exploring Short Gamma-ray Bursts as Gravitational-wave Standard Sirens}},
  doi           = {10.1088/0004-637X/725/1/496},
  eprint        = {0904.1017},
  number        = {1},
  pages         = {496-514},
  volume        = {725},
  adsurl        = {https://ui.adsabs.harvard.edu/abs/2010ApJ...725..496N},
  archiveprefix = {arXiv},
  journal       = {\apj},
  month         = dec,
  primaryclass  = {astro-ph.CO},
  year          = {2010},
}

@Article{barack19,
  author        = {{Barack}, Leor and {Cardoso}, Vitor and {Nissanke}, Samaya and {Sotiriou}, Thomas P. and {Askar}, Abbas and {Belczynski}, Chris and {Bertone}, Gianfranco and {Bon}, Edi and {Blas}, Diego and {Brito}, Richard and {Bulik}, Tomasz and more},
  title         = {{Black holes, gravitational waves and fundamental physics: a roadmap}},
  doi           = {10.1088/1361-6382/ab0587},
  eid           = {143001},
  eprint        = {1806.05195},
  number        = {14},
  pages         = {143001},
  volume        = {36},
  adsurl        = {https://ui.adsabs.harvard.edu/abs/2019CQGra..36n3001B},
  archiveprefix = {arXiv},
  journal       = {Class. Quantum Grav.},
  month         = jul,
  primaryclass  = {gr-qc},
  year          = {2019},
}

@ARTICLE{ridders79,
  author={Ridders, C.},
  journal={IEEE Transactions on Circuits and Systems}, 
  title={A new algorithm for computing a single root of a real continuous function}, 
  year={1979},
  volume={26},
  number={11},
  pages={979-980},
  doi={10.1109/TCS.1979.1084580}}

@ARTICLE{biswas26,
       author = {{Biswas}, B. and {Rosswog}, S. and {Diener}, P. and {Schnabel}, L.},
      journal = {eprint arXiv:2601.01402},
         year = "2026"
}

@ARTICLE{schianchi24,
       author = {{Schianchi}, Federico and {Gieg}, Henrique and {Nedora}, Vsevolod and {Neuweiler}, Anna and {Ujevic}, Maximiliano and {Bulla}, Mattia and {Dietrich}, Tim},
        title = "{M 1 neutrino transport within the numerical-relativistic code BAM with application to low mass binary neutron star mergers}",
      journal = {\prd},
         year = 2024,
        month = feb,
       volume = {109},
       number = {4},
          eid = {044012},
        pages = {044012},
          doi = {10.1103/PhysRevD.109.044012},
archivePrefix = {arXiv},
       eprint = {2307.04572},
 primaryClass = {gr-qc},
       adsurl = {https://ui.adsabs.harvard.edu/abs/2024PhRvD.109d4012S}
}

@Article{gehrels06,
  author        = {{Gehrels}, N. and {Norris}, J.~P. and {Barthelmy}, S.~D. and {Granot}, J. and {Kaneko}, Y. and {Kouveliotou}, C. and {Markwardt}, C.~B. and {M{\'e}sz{\'a}ros}, P. and {Nakar}, E. and {Nousek}, J.~A. and {O'Brien}, P.~T. and {Page}, M. and {Palmer}, D.~M. and {Parsons}, A.~M. and {Roming}, P.~W.~A. and {Sakamoto}, T. and {Sarazin}, C.~L. and {Schady}, P. and {Stamatikos}, M. and {Woosley}, S.~E.},
  title         = {{A new {\ensuremath{\gamma}}-ray burst classification scheme from GRB060614}},
  doi           = {10.1038/nature05376},
  eprint        = {astro-ph/0610635},
  number        = {7122},
  pages         = {1044-1046},
  volume        = {444},
  adsurl        = {https://ui.adsabs.harvard.edu/abs/2006Natur.444.1044G},
  archiveprefix = {arXiv},
  journal       = {Nature},
  month         = dec,
  primaryclass  = {astro-ph},
  year          = {2006},
}

@Article{rosswog22b,
  author        = {{Rosswog}, Stephan and {Diener}, Peter and {Torsello}, Francesco},
  title         = {{Thinking Outside the Box: Numerical Relativity with Particles}},
  doi           = {10.3390/sym14061280},
  eprint        = {2205.08130},
  number        = {6},
  pages         = {1280},
  volume        = {14},
  adsurl        = {https://ui.adsabs.harvard.edu/abs/2022Symm...14.1280R},
  archiveprefix = {arXiv},
  journal       = {Symmetry},
  month         = jun,
  primaryclass  = {gr-qc},
  year          = {2022},
}

@Article{rosswog10a,
  author  = {{Rosswog}, S.},
  title   = {{Relativistic smooth particle hydrodynamics on a given background spacetime}},
  doi     = {10.1088/0264-9381/27/11/114108},
  number  = {11},
  pages   = {114108},
  volume  = {27},
  adsurl  = {http://adsabs.harvard.edu/abs/2010CQGra..27k4108R},
  journal = {Class. Quantum Grav.},
  month   = jun,
  year    = {2010},
}

@Article{kiuchi15,
  author        = {{Kiuchi}, K. and {Cerd{\'a}-Dur{\'a}n}, P. and {Kyutoku}, K. and {Sekiguchi}, Y. and {Shibata}, M.},
  title         = {{Efficient magnetic-field amplification due to the Kelvin-Helmholtz instability in binary neutron star mergers}},
  doi           = {10.1103/PhysRevD.92.124034},
  eid           = {124034},
  eprint        = {1509.09205},
  number        = {12},
  pages         = {124034},
  volume        = {92},
  adsurl        = {https://ui.adsabs.harvard.edu/abs/2015PhRvD..92l4034K},
  archiveprefix = {arXiv},
  journal       = {\prd},
  month         = dec,
  primaryclass  = {astro-ph.HE},
  year          = {2015},
}

@ARTICLE{kastaun21,
       author = {{Kastaun}, Wolfgang and {Kalinani}, Jay Vijay and {Ciolfi}, Riccardo},
        title = "{Robust recovery of primitive variables in relativistic ideal magnetohydrodynamics}",
      journal = {\prd},
         year = 2021,
        month = jan,
       volume = {103},
       number = {2},
          eid = {023018},
        pages = {023018},
          doi = {10.1103/PhysRevD.103.023018},
archivePrefix = {arXiv},
       eprint = {2005.01821},
 primaryClass = {gr-qc},
       adsurl = {https://ui.adsabs.harvard.edu/abs/2021PhRvD.103b3018K}
}

@ARTICLE{guo25,
       author = {{Guo}, Gang and {Qian}, Yong-Zhong and {Wu}, Meng-Ru},
        title = "{Binary neutron star mergers as potential sources for ultrahigh-energy cosmic rays and high-energy neutrinos}",
      journal = {\prd},
         year = 2025,
        month = sep,
       volume = {112},
       number = {6},
          eid = {063022},
        pages = {063022},
          doi = {10.1103/zfp3-y9yw},
archivePrefix = {arXiv},
       eprint = {2506.17581},
 primaryClass = {astro-ph.HE},
       adsurl = {https://ui.adsabs.harvard.edu/abs/2025PhRvD.112f3022G}
}

@ARTICLE{cerda_duran08,
       author = {{Cerd{\'a}-Dur{\'a}n}, P. and {Font}, J.~A. and {Ant{\'o}n}, L. and {M{\"u}ller}, E.},
        title = "{A new general relativistic magnetohydrodynamics code for dynamical spacetimes}",
      journal = {\aap},
         year = 2008,
        month = dec,
       volume = {492},
       number = {3},
        pages = {937-953},
          doi = {10.1051/0004-6361:200810086},
archivePrefix = {arXiv},
       eprint = {0804.4572},
 primaryClass = {astro-ph},
       adsurl = {https://ui.adsabs.harvard.edu/abs/2008A&A...492..937C}
}

@Article{anderson08b,
  author        = {{Anderson}, M. and {Hirschmann}, E.~W. and {Lehner}, L. and {Liebling}, S.~L. and {Motl}, P.~M. and {Neilsen}, D. and {Palenzuela}, C. and {Tohline}, J.~E.},
  title         = {{Magnetized Neutron-Star Mergers and Gravitational-Wave Signals}},
  doi           = {10.1103/PhysRevLett.100.191101},
  eprint        = {0801.4387},
  number        = {19},
  pages         = {191101},
  volume        = {100},
  adsurl        = {http://adsabs.harvard.edu/abs/2008PhRvL.100s1101A},
  archiveprefix = {arXiv},
  journal       = {\prl},
  month         = may,
  primaryclass  = {gr-qc},
  year          = {2008},
}

@Book{press92,
  author    = {W. H. Press and B. P. Flannery and S. A. Teukolsky and W. T. Vetterling},
  title     = {Numerical Recipes},
  publisher = {Cambridge University Press},
  address   = {New York},
  year      = {1992},
}

@Article{kasen17,
  author        = {{Kasen}, D. and {Metzger}, B. and {Barnes}, J. and {Quataert}, E. and {Ramirez-Ruiz}, E.},
  title         = {{Origin of the heavy elements in binary neutron-star mergers from a gravitational-wave event}},
  doi           = {10.1038/nature24453},
  eprint        = {1710.05463},
  pages         = {80-84},
  volume        = {551},
  adsurl        = {http://adsabs.harvard.edu/abs/2017Natur.551...80K},
  archiveprefix = {arXiv},
  journal       = {Nature},
  month         = nov,
  primaryclass  = {astro-ph.HE},
  year          = {2017},
}

@Article{schutz86,
  author  = {Schutz, B.F.},
  title   = {{Determining the {H}ubble constant from gravitational wave observations}},
  pages   = {310},
  volume  = {323},
  journal = {Nature},
  year    = {1986},
}

@Article{kiuchi24,
  author        = {{Kiuchi}, Kenta and {Reboul-Salze}, Alexis and {Shibata}, Masaru and {Sekiguchi}, Yuichiro},
  title         = {{A large-scale magnetic field produced by a solar-like dynamo in binary neutron star mergers}},
  doi           = {10.1038/s41550-024-02194-y},
  eprint        = {2306.15721},
  pages         = {298-307},
  volume        = {8},
  adsurl        = {https://ui.adsabs.harvard.edu/abs/2024NatAs...8..298K},
  archiveprefix = {arXiv},
  journal       = {Nature Astronomy},
  month         = mar,
  primaryclass  = {astro-ph.HE},
  year          = {2024},
}

@ARTICLE{rossoni25,
       author = {{Rossoni}, Simone and {Sigl}, G{\"u}nter},
        title = "{Anisotropy signal of ultrahigh-energy cosmic rays from a structured magnetized universe}",
      journal = {\prd},
         year = 2025,
        month = jul,
       volume = {112},
       number = {2},
          eid = {023015},
        pages = {023015},
          doi = {10.1103/4zt7-s48v},
       adsurl = {https://ui.adsabs.harvard.edu/abs/2025PhRvD.112b3015R}
}

@Article{rosswog23a,
  author        = {{Rosswog}, Stephan and {Torsello}, Francesco and {Diener}, Peter},
  title         = {{The Lagrangian Numerical Relativity code SPHINCS\_BSSN\_v1.0}},
  doi           = {10.48550/arXiv.2306.06226},
  eprint        = {2306.06226},
  volume        = {9},
  adsurl        = {https://ui.adsabs.harvard.edu/abs/2023arXiv230606226R},
  archiveprefix = {arXiv},
  journal       = {Front. Appl. Math. Stat.},
  month         = jun,
  primaryclass  = {gr-qc},
  year          = {2023},
}

@Article{rosswog20a,
  author        = {{Rosswog}, S.},
  title         = {{The Lagrangian hydrodynamics code MAGMA2}},
  doi           = {10.1093/mnras/staa2591},
  eprint        = {1911.13093},
  number        = {3},
  pages         = {4230-4255},
  volume        = {498},
  adsurl        = {https://ui.adsabs.harvard.edu/abs/2020MNRAS.498.4230R},
  archiveprefix = {arXiv},
  journal       = {\mnras},
  month         = {aug},
  primaryclass  = {astro-ph.IM},
  year          = {2020},
}

@Article{troja22,
  author        = {{Troja}, E. and {van Eerten}, H. and {Zhang}, B. and {Ryan}, G. and {Piro}, L. and {Ricci}, R. and {O'Connor}, B. and {Wieringa}, M.~H. and {Cenko}, S.~B. and {Sakamoto}, T.},
  title         = {{A thousand days after the merger: Continued X-ray emission from GW170817}},
  doi           = {10.1093/mnras/staa2626},
  eprint        = {2006.01150},
  number        = {4},
  pages         = {5643-5651},
  volume        = {498},
  adsurl        = {https://ui.adsabs.harvard.edu/abs/2020MNRAS.498.5643T},
  archiveprefix = {arXiv},
  journal       = {\mnras},
  month         = nov,
  primaryclass  = {astro-ph.HE},
  year          = {2020},
}

@Article{sekiguchi16a,
       author = {{Sekiguchi}, Yuichiro and {Kiuchi}, Kenta and {Kyutoku}, Koutarou and {Shibata}, Masaru and {Taniguchi}, Keisuke},
        title = "{Dynamical mass ejection from the merger of asymmetric binary neutron stars: Radiation-hydrodynamics study in general relativity}",
      journal = {\prd},
         year = 2016,
        month = jun,
       volume = {93},
       number = {12},
          eid = {124046},
        pages = {124046},
          doi = {10.1103/PhysRevD.93.124046},
archivePrefix = {arXiv},
       eprint = {1603.01918},
 primaryClass = {astro-ph.HE},
       adsurl = {https://ui.adsabs.harvard.edu/abs/2016PhRvD..93l4046S}
}

@Article{rosswog21a,
  author        = {{Rosswog}, S. and {Diener}, P.},
  title         = {{SPHINCS\_BSSN: a general relativistic smooth particle hydrodynamics code for dynamical spacetimes}},
  doi           = {10.1088/1361-6382/abee65},
  eid           = {115002},
  eprint        = {2012.13954},
  number        = {11},
  pages         = {115002},
  volume        = {38},
  adsurl        = {https://ui.adsabs.harvard.edu/abs/2021CQGra..38k5002R},
  archiveprefix = {arXiv},
  journal       = {Class. Quantum Grav.},
  month         = jun,
  primaryclass  = {gr-qc},
  year          = {2021},
}

@Book{rezzolla13a,
  author    = {Rezzolla, L. and Zanotti, O.},
  title     = {Relativistic Hydrodynamics},
  doi       = {10.1093/acprof:oso/9780198528906.001.0001},
  publisher = {Oxford University Press},
  address   = {Oxford; New York},
  adsurl    = {http://adsabs.harvard.edu/abs/2013rehy.book.....R},
  year      = {2013},
}

@article{Shankar_2023,
doi = {10.1088/1361-6382/acf2d9},
url = {https://doi.org/10.1088/1361-6382/acf2d9},
year = {2023},
month = {sep},
publisher = {IOP Publishing},
volume = {40},
number = {20},
pages = {205009},
author = {Shankar, Swapnil and Mösta, Philipp and Brandt, Steven R and Haas, Roland and Schnetter, Erik and de Graaf, Yannick},
title = {GRaM-X: a new GPU-accelerated dynamical spacetime GRMHD code for Exascale computing with the Einstein Toolkit},
journal = {Classical and Quantum Gravity}
}

@article{Siegel_2018,
doi = {10.3847/1538-4357/aabcc5},
url = {https://doi.org/10.3847/1538-4357/aabcc5},
year = {2018},
month = {may},
publisher = {The American Astronomical Society},
volume = {859},
number = {1},
pages = {71},
author = {Siegel, Daniel M. and Mösta, Philipp and Desai, Dhruv and Wu, Samantha},
title = {Recovery Schemes for Primitive Variables in General-relativistic Magnetohydrodynamics},
journal = {The Astrophysical Journal}
}

@article{Cupp_2026,
  title = {Modern, infrastructure-agnostic, extensible library for GRMHD simulations},
  author = {Cupp, Samuel and Werneck, Leonardo R. and Jacques, Terrence Pierre and Tootle, Samuel and Etienne, Zachariah B.},
  journal = {Phys. Rev. D},
  volume = {113},
  issue = {4},
  pages = {043045},
  numpages = {24},
  year = {2026},
  month = {Feb},
  publisher = {American Physical Society},
  doi = {10.1103/bbtm-31r5},
  url = {https://link.aps.org/doi/10.1103/bbtm-31r5}
}

@article{Noble_2006,
doi = {10.1086/500349},
url = {https://doi.org/10.1086/500349},
year = {2006},
month = {apr},
publisher = {},
volume = {641},
number = {1},
pages = {626},
author = {Noble, Scott C. and Gammie, Charles F. and McKinney, Jonathan C. and Del Zanna, Luca},
title = {Primitive Variable Solvers for Conservative General Relativistic Magnetohydrodynamics},
journal = {The Astrophysical Journal}
}

@misc{rosswog2026,
      title={SPH methods in the modelling of compact objects}, 
      author={Stephan Rosswog},
      year={2026},
      eprint={2607.14828},
      archivePrefix={arXiv},
      primaryClass={astro-ph.HE},
      url={https://arxiv.org/abs/2607.14828}, 
}

@article{Banyuls_1997,
doi = {10.1086/303604},
url = {https://doi.org/10.1086/303604},
year = {1997},
month = {feb},
publisher = {},
volume = {476},
number = {1},
pages = {221},
author = {Banyuls, Francesc and Font, José A. and Ibáñez, José Ma. and Martí, José Ma. and Miralles, Juan A.},
title = {Numerical {3 + 1} General Relativistic Hydrodynamics: A Local Characteristic Approach},
journal = {The Astrophysical Journal}
}

@article{Anton_2006,
doi = {10.1086/498238},
url = {https://doi.org/10.1086/498238},
year = {2006},
month = {jan},
publisher = {},
volume = {637},
number = {1},
pages = {296},
author = {Antón, Luis and Zanotti, Olindo and Miralles, Juan A. and Martí, José M. and Ibáñez, José M. and Font, José A. and Pons, José A.},
title = {Numerical 3+1 General Relativistic Magnetohydrodynamics: A Local Characteristic Approach},
journal = {The Astrophysical Journal}
}
\bibliographystyle{aasjournalv7}



\end{document}